\documentclass[review]{elsarticle}

\usepackage[a4paper, margin=1in]{geometry}
\usepackage{setspace}
\spacing{1.2}
\usepackage{xcolor}
\usepackage{soul}
\usepackage{mathtools}
\usepackage{nicefrac}
\usepackage{textcomp, gensymb}
\usepackage{amsmath}
\usepackage{lipsum,times,graphicx,hyperref,cleveref}
\usepackage{multicol}
\usepackage{caption}

\newenvironment{nomenclature}[1][1em]
 {\begin{list}{}{%
   \settowidth{\labelwidth}{\textbf{#1}}%
   \setlength{\leftmargin}{\labelwidth}%
   \addtolength{\leftmargin}{\labelsep}%
 }}
 {\end{list}}

\newcommand{\nomenclatureentry}[2]{\item[#1] #2\par}

\begin{document}

    \begin{sloppypar}

    \begin{frontmatter}

    \title{\textbf{Response of premixed Jet Flames to Blast waves}}
    
    \author[mymainaddress]{Akhil Aravind}
    
    \author[mymainaddress]{Gautham Vadlamudi}
    
    \author[mymainaddress,mysecondaryaddress]{Saptarshi Basu\corref{mycorrespondingauthor}}
    \cortext[mycorrespondingauthor]{Corresponding author: sbasu@iisc.ac.in}
    
    \address[mymainaddress]{Department of Mechanical Engineering, Indian Institute of Science, Bangalore}
    \address[mysecondaryaddress]{Interdisciplinary Center for Energy Research, Indian Institute of Science, Bangalore}
    
    \begin{abstract}
    The study investigates the response dynamics of premixed jet flames when incident with blast waves along the jet axis. In the present work, blast waves are generated using the wire-explosion technique.
    The generated blast wave interacts with premixed jet flames that are stabilised over a thin fuel-air jet nozzle. The study is performed over a wide range of parametric space, varying the Reynolds number ($Re$) and normalised equivalence ratio ($\Phi$) of the premixed jet and the strength of the generated blast fronts (characterized by $M_{s,r}$). 
    The blast wave imposes a decaying flow field profile characterised by a sharp discontinuity at the blast front. This is accompanied by a subsequent induced bulk flow arising due to entertainment from the surroundings as the blast-imposed flow fields decay to sub-ambient levels. 
    While the jet flame is observed to respond to the blast front with a jittery motion, it is found to lift off following the interaction with the induced flow. Depending on the operating conditions ($Re$, $\Phi$, and $M_{s,r}$), the flame lift-off is followed by an extinction or a re-attachment event. The flame response is further classified into two re-attachment and three extinction sub-regimes based on the response dynamics of the flame base and flame tip following the interaction process. 
    The flame response entails flame base lift-off and flame tip stretching/distortion, potentially leading to a flame pinch-off event contingent on the operating conditions. A simplified mathematical model was developed to explain the flame base response dynamics, yielding a scaling law for flame base lift-off height. Flame tip response trends were elucidated by extending the vorticity transport equation to estimate the vortex roll-up rate in the shear boundary surrounding the flame. Flame pinch-off was observed when shear layer vortices reached critical circulation limits and shed at length scales lower than the flame height.
    \end{abstract}
    
    
    \end{frontmatter}

    \section*{Novelty and Significance Statement}
    {
    This work is the first of its kind to examine the response of premixed jet flames to blast waves that are incident along the direction of the jet axis. The study identifies qualitative and quantitative changes induced in the premixed jet flame following the interaction process and proposes a regime map classifying the flame response behaviour across the parametric space of premixed jet Reynolds number, equivalence ratio, and strength of the incident blast wave. The axial blast-flame interaction setting explored in the current work can provide valuable insights into comprehending and improving practical high-speed propulsion systems such as scram jets and RDEs, wherein flame interactions with non-linear pressure waves, such as decaying shocks and blast waves, are common. 
    }
    
    \section*{Author Contributions}
    {
    \textbf{Akhil Aravind:} Conceptualisation, Experiments, Data Analysis, Writing - Original Draft. \textbf{Gautham Vadlamudi:} Conceptualisation, Experiments, Data Analysis, Writing - Review \& Editing. \textbf{Saptarshi Basu:} Conceptualisation, Data Analysis, Writing - Review \& Editing, Fund acquisition.
    }

    \newpage

    \section*{Nomenclature}
    
    \begin{spacing}{1}

    \begin{multicols}{2}
    \begin{nomenclature}
      \nomenclatureentry{$Re$}{Premixed jet Reynolds Number}
      \nomenclatureentry{$M_{s}$}{Blast Mach Number}
      \nomenclatureentry{$M_{s}$}{Blast Radius}
      \nomenclatureentry{$M_{s,r}$}{Reference Blast Mach Number}
      \nomenclatureentry{$\phi$}{Equivalence ratio}
      \nomenclatureentry{$\Phi$}{Normalised equivalence ratio}  
      \nomenclatureentry{$d$}{Inner diameter of the central tube}
      \nomenclatureentry{$h$}{Flame height}
      \nomenclatureentry{$v_{s}$}{Blast imposed velocity field}
      \nomenclatureentry{$v_{p}$}{Velocity peak imposed by the blast wave}
      \nomenclatureentry{$p_{s}$}{Blast imposed pressure field}
      \nomenclatureentry{$p_{p}$}{Pressure peak imposed by the blast wave}
      \nomenclatureentry{$t_{i}$}{Instance corresponding to the incidence of blast with flame}
      \nomenclatureentry{$t_{s,v}$}{Instance corresponding decay of $v_{s}$ to ambient levels}
      \nomenclatureentry{$t_{s,p}$}{Instance corresponding decay of $p_{s}$ to ambient levels}
      \nomenclatureentry{$t_{0}$}{Instance corresponding to flame base lift-off}
      \nomenclatureentry{$x_{b}$}{Flame base location}
      \nomenclatureentry{$v_{in}$}{Bulk flow induced behind the blast wave}
      \nomenclatureentry{$S_{L,b}$}{Effective flame speed at the flame base}
      \nomenclatureentry{$h_{b,lft}$}{Flame base lift-off height}
      \nomenclatureentry{$t_{b,lft}$}{Timescale associated with flame base lift-off}
      \nomenclatureentry{$t_{ra}$}{Re-attachment time associated with the flame base}
      \nomenclatureentry{$f_{sh}$}{Flame tip shedding frequency}
      \nomenclatureentry{$h_{sh}$}{Flame tip shedding height}
      \nomenclatureentry{$t_{sh}$}{Flame tip shedding time period}
      \nomenclatureentry{$t_{b,ext}$}{Timescale associated with flame extinction}
      \nomenclatureentry{$d_{b,f}$}{Diameter of the flame base}
      \nomenclatureentry{$d_{o}$}{Diameter of the outer tube in the co-axial jet analogy}
      \nomenclatureentry{$\rho_{j}$}{Density of the premixed fuel-air jet}
      \nomenclatureentry{$\rho_{a}$}{Density of ambient air}
      \nomenclatureentry{$v_{f}$}{Effective velocity of the premixture jet in the co-axial jet analogy}
      \nomenclatureentry{$Re_{f}$}{Effective Reynolds number of the premixture jet in the co-axial jet analogy}
      \nomenclatureentry{$\phi_{b}$}{Effective equivalence ratio of the premixture jet in the co-axial jet analogy}
      \nomenclatureentry{$D$}{Binary diffusion coefficient}
      \nomenclatureentry{$\Gamma$}{Circulation}
      \nomenclatureentry{$v_{j}$}{Fuel-air jet velocity}
      \nomenclatureentry{$v_{nc}$}{Velocity scale of natural convection}
    \end{nomenclature}
    \end{multicols}

    \end{spacing}

    \section{Introduction}

    Understanding shock flame interactions is imperative due to their widespread implications across various fields. These interactions are known to cause flame distortion, enhance mixing, accelerate the flame to supersonic conditions \cite{thomas_experimental_2001}, and, in extreme cases, destabilise the flame and cause blow-off \cite{yoshida_blowoff_2024}. One significant area where this phenomenon is prominent is in high-speed propulsion systems such as scramjets, where shock-flame interactions significantly shape the overall combustion dynamics \cite{yang_applications_1993}. Additionally, understanding these interactions is central to comprehending processes like deflagration to detonation transition (DDT) \cite{oran_origins_2007}, which are crucial in futuristic propulsion systems like rotating detonation engines (RDE).
    
    Furthermore, beyond propulsion systems, the study of shock flame interactions holds pivotal importance in comprehending and improving large-scale fire extinguishing systems employing explosives/shock-waves to blow off flames. Such systems employ shocks (or blast waves from explosions) to drive away the existing flame from the fuel source and have proven to be an effective method in fighting wildfires and oil fires \cite{akhmetov_extinguishing_1980,akhmetov_formation_2001,yoshida_blowoff_2024}. Thus, delving into the complexities of shock flame interactions not only advances our understanding of propulsion systems but also contributes significantly to enhancing fire safety measures.
    
    Markstein et al. \cite{markstein_shock-tube_1957} pioneered the shock-flame interaction studies, investigating the distortion of premixed butane-air flames upon interacting them with shock waves at different Mach numbers. He observed the reversal of the original flame shape and the formation of an unburned gas funnel into the product gases as the shock swept across the flame from the unburnt end to the burnt end. Subsequent studies by Rudinger et al. \cite{rudinger_shock_1958} confirmed this behaviour, and they showed that the observed funnel of unburnt gases would eventually evolve into a vortex ring at later time scales. Wei et al. \cite{wei_effect_2017} further demonstrated that such flame distortion events following the interaction process would result in local enhancement in heat release rates and temperatures. Khokhlov et al. \cite{khokhlov_interaction_1999} identified Richtmyer Meshkov (RM) instability as the prominent factor driving such shock-induced flame distortion and highlighted the need for multiple such interactions to achieve significant reaction rate enhancements and flame acceleration to supersonic speeds. Studies from Thomas et al. \cite{thomas_experimental_2001} involving a transverse interaction of a laminar flame bubble with a shock front revealed that the associated rise in local temperature and pressure following the multiple shock-flame interactions processes transitions the deflagration wave into a detonation front. Numerical studies by Attal et al. \cite{attal_numerical_2015} explored the complex post-interaction dynamics, revealing the influence of non-linearity and Rayleigh-Taylor instability on the growth rate of flame interface perturbations at timescales, which are responsible for significant flame distortion events.
    
    In practical situations, unlike controlled shocks generated by shock tubes, shockwaves typically arise when blast waves converge or focus due to the geometric features of the flow domain, leading to the formation of unsteady shocks. This phenomenon results in a flow characterised by a sharp discontinuity, followed by a decaying flow field that falls below ambient levels, leading to entrainment from the surroundings. Expanding on this concept, Fleche et al. \cite{la_fleche_dynamics_2018} investigated the dynamics of head-on interaction between blast waves and cellular flames in Hele-Shaw cells. Their unique experimental facility allowed them to explore the response dynamics of cellular flames when a blast wave sweeps over the flame from the unburnt zone to the burnt zone and vice versa. While the flame exhibited significant distortion and reversal of the cellular flame front immediately following the interaction with the blast front in the former case, the latter case showed competing flame distortion effects from RM and RT instability, wherein the RT mode eventually overpowered and caused a reversal of the density interface between the unburnt and burnt gases. 
    
    Exploring the dynamics of shock-flame interactions in more practical flame configurations, Chan et al. \cite{doig_interaction_2013,doig_shock_2012,chan_interactions_2016} investigated the response of non-premixed flames (from small-scale Bunsen burners to large-scale ring burners) to the transverse incidence of high-speed exhaust flows from a shock tube. Their study characterised the exhaust flow field as a planar shock front, followed by a high-impulse flow (blast wind) and a vortex ring formed by entrained rotating flows. While the passage of the shock front showed no noticeable flame disruption, the flow features following it were found to extinguish the flame. Their studies characterised the minimum transverse flow velocity required to extinguish the flame. 
    
    Extending upon the above-mentioned studies, the present study explores the response dynamics of premixed jet flames when incident with blast waves along the flame-jet axis. The experimental facility employed in the current study relies on the wire explosion technique to generate blast waves. The technique was demonstrated by Oshima et al. \cite{oshima_blast_1962}, wherein a fine metal wire is exploded by imposing a high voltage electrical pulse across it, producing a cylindrical blast front along its length. The technique was more recently employed by Sharma et al. \cite{sharma_shock_2021,sharma_shock-induced_2023} and Chandra et al. \cite{chandra_shock-induced_2023} to study the shock-induced secondary atomisation of Newtonian and non-Newtonian fluid droplets, respectively. Chandra et al. \cite{chandra_shock-induced_2023} highlighted the characteristic sharp decay in the blast-imposed flow field profiles following the discontinuity at the blast front. As blast-imposed pressure decays and drops below ambient conditions in a characteristic decay time scale, entertainment from the surroundings is expected, causing a second spike in the flow field profiles. The associated large spatiotemporal flow field gradients are expected to alter the response dynamics of the premixed jet flame significantly. It is also interesting to note that, regardless of the source geometry, the generated blast waves tend towards spherical symmetry at large propagation distances (relative to the source size) \cite{almustafa_fundamental_2023}.
    
    In addition to flame distortion, the axial interaction between blast waves and jet flames is poised to modify the canonical characteristics of the jet flame. Jet flames exhibit a distinct flickering pattern, which is attributed to the buoyancy-induced shear layer roll-up along the density interface between the hot product gases and the surrounding ambient fluid. Tip flickering corresponds to the periodic vortex shedding events that occur as the shear layer vortices grow to reach their critical circulation limits and detach from their feeding interface shear layer \cite{xia_vortex-dynamical_2018}. A significant number of studies \cite{chamberlin_flicker_1948,cetegen_experiments_1993} over the past decade have contributed to our understanding of flame tip flickering. For buoyancy-dominated jet flames, wherein the Froude number (Froude number, $Fr$, characterises the relative importance of inertial forces against buoyant forces acting on the flame) tends to zero, the non-dimensional flickering frequency (characterised by the Strouhal number) has been found to scale with $Fr^{-0.5}$ \cite{cetegen_experiments_1993}. A similar scaling law was proposed by Hamins et al. \cite{hamins_experimental_1992} at high Froude numbers ($10^{-4} < Fr < 10^{8}$), alongside identifying the critical fuel jet velocity limits to initiate flame flickering. Studies by Thirumalaikumaran et al. \cite{thirumalaikumaran_insight_2022} and Pandey et al. \cite{pandey_dynamic_2021} have demonstrated that external flows imposed over diffusion flames have significant effects on their flickering characterises. External flows are found to accelerate the roll-up rate of shear layer vortices, leading to more frequent vortex shedding events (at shorter time scales) and a decrease in the observed flame flickering height (flame height at which shear layer vortices shed). Extending from the above-mentioned studies, a change in the dynamics of tip flickering is expected following the interaction of the jet flames with blast waves. 
    
    Thus, the present study aims to comprehend flame distortion, flame tip flickering, and stability characteristics of premixed jet flames post axial interactions with blast waves and assess the impact of blast wave Mach numbers, fuel-air mixture velocities, and equivalence ratio on flame’s response dynamics.

    \section{Experimental Setup}

    \begin{figure}[h]
        \centering
        \includegraphics[width=1\linewidth]{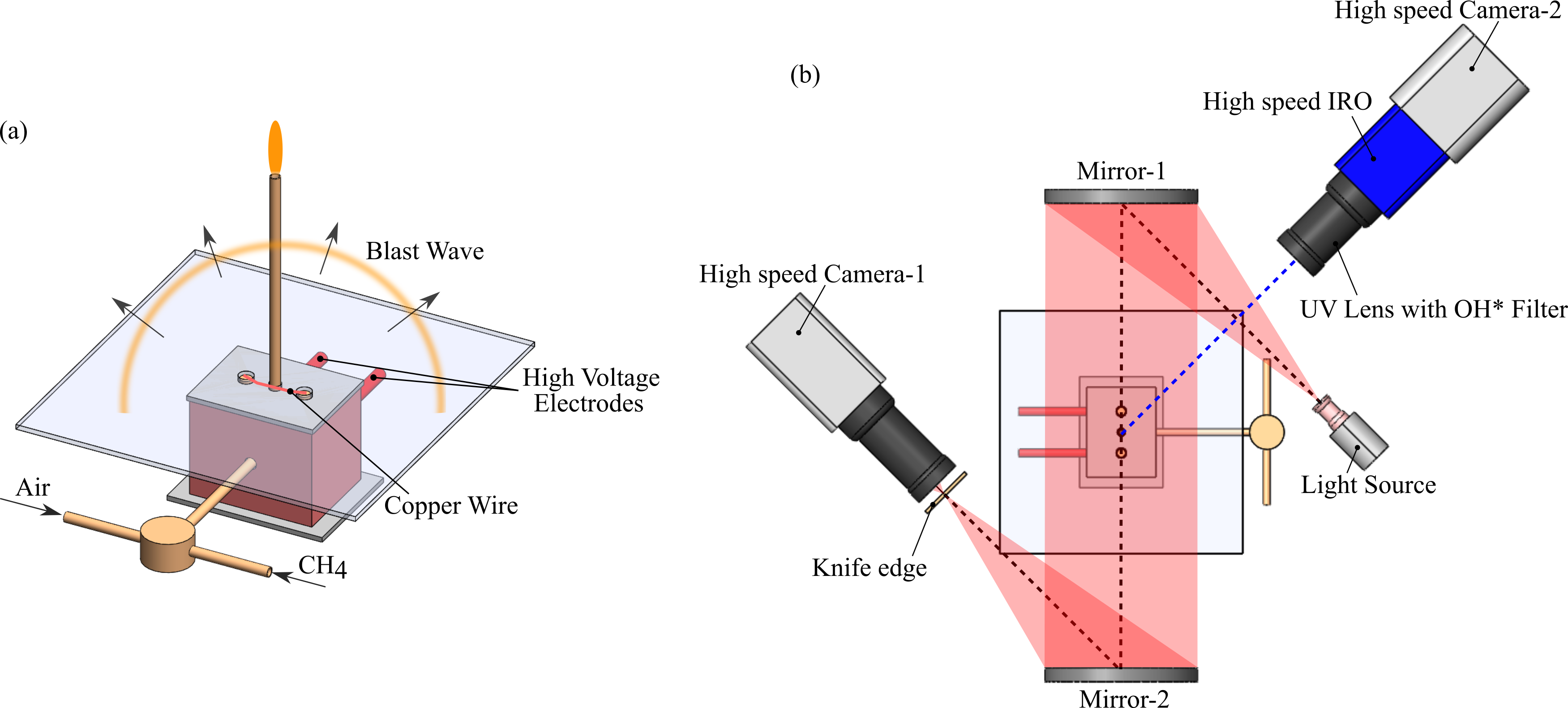}
        \caption{(a) Schematic of the high-voltage electrode chamber that is used for blast wave generation. A stainless steel tube passes through the chamber's base plate and stabilizes the premixed jet flame at a height of 264mm. (b) Schematic depicting the high-speed recording systems (for Schlieren flow visualization and OH* chemiluminescence) used to study the interaction process}
        \label{fig:Exp_setup}
    \end{figure}
    
    The experimental facility employed in the current work is schematically depicted in \textbf{Fig. \ref{fig:Exp_setup}(a)}. The facility consists of an electrode chamber for producing the shock/blast waves and works on the principle of high-voltage electrical wire explosion \cite{oshima_blast_1962}. The system comprises a power source (Zeonics Systech, India Z/46/12) housing a 5 $\mu$F capacitor, which can be charged to desired energy levels through a charging circuit running through the main power supply. Once charged, the charging circuit is cut off, and the capacitor is only allowed to discharge through a pair of electrodes that are housed in the electrode chamber (forming the discharge circuit). A thin copper wire of size 35 SWG establishes electrical contact between the high-voltage electrodes (\textbf{Fig. \ref{fig:Exp_setup}}(a)). The discharge circuit is triggered (brought into electrical contact with the charged capacitor) upon receiving a TTL pulse from an external delay generator that syncs the blast generator with the high-speed recording systems (high-speed cameras used for Schlieren flow visualization and OH chemiluminescence). As the capacitor discharges through the thin copper wire (connected across the electrodes), rapid joule's heating causes the wire to vaporize and produce a cylindrical blast wave along its length \cite{oshima_blast_1962}. 
    
    Increasing the charging voltage was found to produce blast waves at higher Mach numbers, and a plot depicting their dependence is provided in supplementary Figure \textbf{S1}. In the current work, the charging voltage was adjusted in increments of 1kV, ranging from 4kV to 7kV. This variation resulted in corresponding blast wave Mach numbers of 1.025, 1.040, 1.060, and 1.075 (as measured at a reference distance of 264mm from the electrode base plate), respectively, all of which are close to the acoustic limit. In the sections that follow, these reference Mach numbers ($M_{s,r}$) are used to refer to the cases corresponding to their respective charging voltages. 
    
    \begin{figure}[h]
        \centering
        \includegraphics[width=0.9\linewidth]{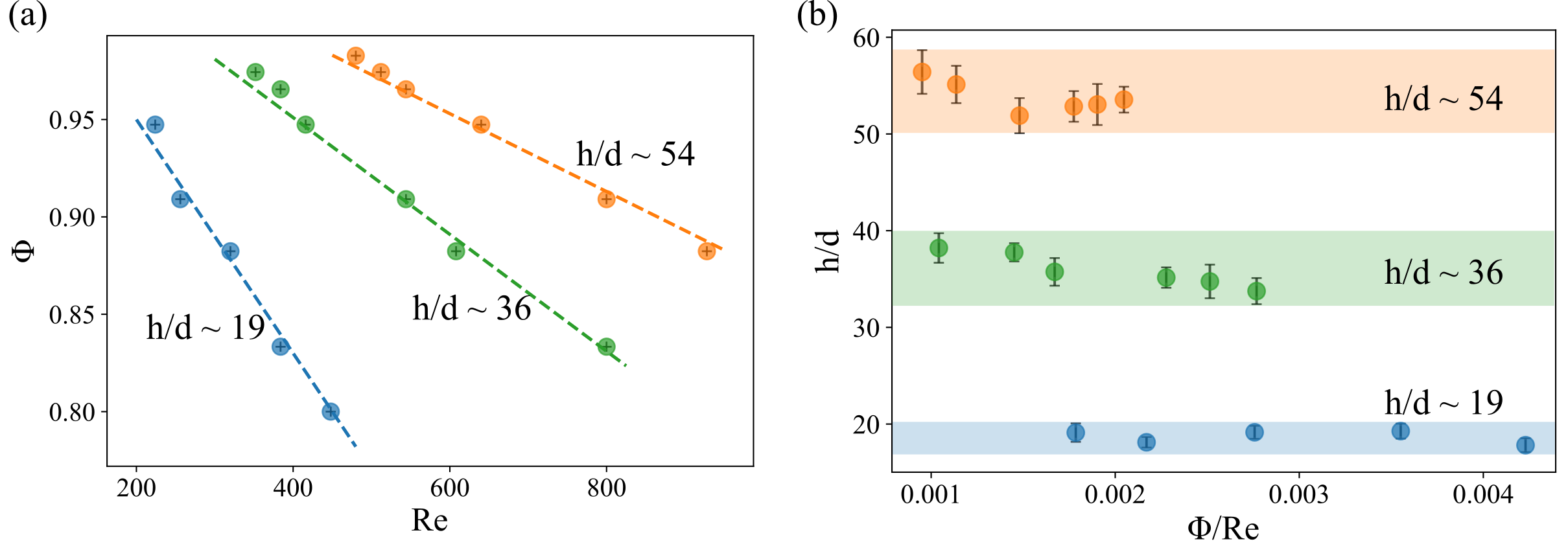}
        \caption{(a) Explored parametric space of the premixed jet flame in terms of $Re$ and $\Phi$. (b) Normalised flame height as a function of $\Phi/Re$}
        \label{fig:para_space}
    \end{figure}
    A central stainless steel tube of 2mm inner diameter ($d$) (that acts as a flame jet nozzle) passes through the centre of the electrode base plate (that supports the copper wire in between the electrodes) and stabilizes the premixed jet flame at a distance of 264 mm from the base plate for axial interaction with the generated blast wave (depicted in \textbf{ Fig. \ref{fig:Exp_setup}}(a)). Fuel (methane) and air are fed through two precise mass flow controllers (Bronkhorst Flexi-Flow Compact with a range of 0–1.6 SLPM for $CH_4$ and 0–2 SLPM for air) into a mixing chamber. The mixture is then directed into the central stainless steel tube. The fuel flow rate was varied from 0.2 SLPM to 0.6 SLPM in steps of 0.2 SLPM. Corresponding to each case, the air flow rate was gradually increased in steps of 0.05 SLPM till the premixed jet flame could no longer be stabilized on the tube rim due to blow-off. Consequently, an increment in the premixed jet velocity was accompanied by a reduction in the equivalence ratio. The resulting parametric space explored in the current work, in terms of premixed jet Reynolds number ($Re$) and normalised equivalence ratio ($\Phi$), is plotted in \textbf{Fig. \ref{fig:para_space} (a)}. It is interesting to note that the premixed flame height ($h$) was found to remain at a near-constant level (variations close to $10\%$) for a given fuel flow rate. The observed variations in $h/d$ are depicted in \textbf{Fig. \ref{fig:para_space} (b)}.
    
    Thus, the experiments were carried out at three different flame height conditions ($h/d \sim$ 19, 36 and 54), corresponding to three different fuel flow rates. The curves representing near-constant flame heights in \textbf{Fig. \ref{fig:para_space} (a)} were found to exhibit a decreasing trend in the magnitude of their slope $\Bigl(\Big|\frac{\triangle \Phi}{\triangle Re}\Big|\Bigr)$ with increasing flame heights. 
    
    During the experimental runs, the premixed fuel-air jet is ignited using a pilot flame and stabilized at the tip of the central tube. The external delay generator then triggers the shock generator and the high-speed recording systems to study the axial interaction between the jet flame and the blast wave.
    
    Flow visualization is achieved through a Schlieren setup employing two parabolic concave mirrors and a high-speed non-coherent pulse diode laser with a wavelength of 640nm (Cavitar Cavilux smart UHS, 400 W power) acting as the light source (\textbf{Fig. \ref{fig:Exp_setup}(b)}). A Photron FASTCAM SA5 camera was used to record the high-speed schlieren data at 40000 fps. Simultaneously, OH* chemiluminescence of the flame was captured at 10000 fps using another high-speed camera (Lavision SA5), coupling it with the high-speed intensified relay optics (Lavision HS IRO; IV Generation) alongside a Nikon Rayfact PF10445MF-UV lens and an OH* band-pass filter (310 nm). The spatial resolutions of flame imaging and Schlieren flow visualization were 5.012 px/mm and 3.430 px/mm, respectively.
    
    \subsection{Data processing and shock-jet flame characterization}
    {
    
    \begin{figure}[h]
        \centering
        \includegraphics[width=1\linewidth]{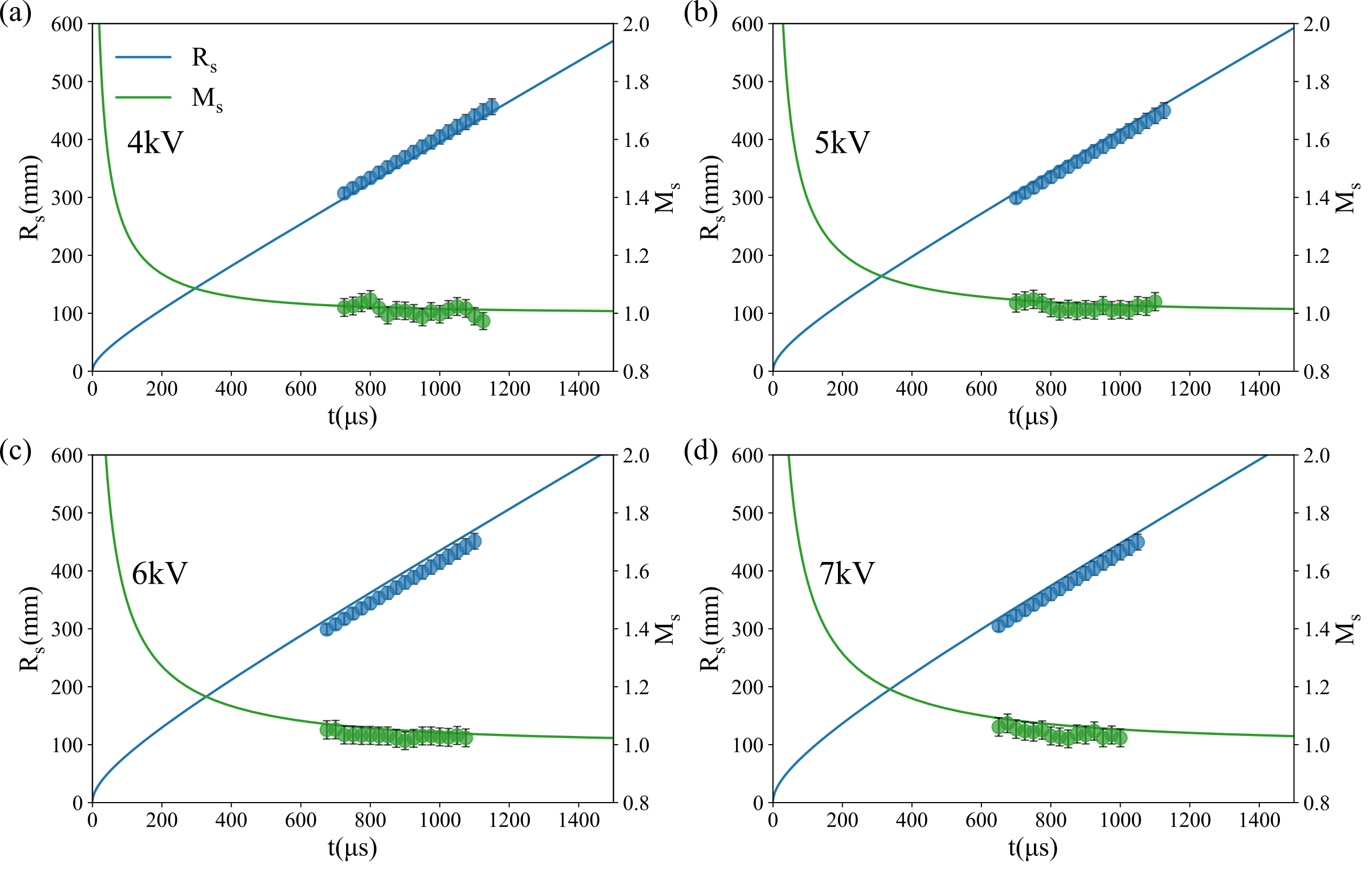}
        \caption{(a-d) Experimentally observed blast wave Mach numbers ($M_{s}$) and radius ($R_{s}$) are compared against the analytical solution proposed by Bach et al. \cite{bach_analytical_1970}. Panels (a-d) correspond to electrode charging voltages of 4kV to 7kV, respectively.}
        \label{fig:expt_bach}
    \end{figure}
    
    The recorded OH* chemiluminescence images of the flame were processed in ImageJ. The Otsu thresholding technique was employed to estimate the flame boundary. The technique is an integral feature of ImageJ, wherein all the pixels in an image are categorized into foreground or background based on an intensity threshold, $I_{f}$, that is estimated by minimizing the variance within each category or maximizing the variance between the two categories. Pixels with intensities greater than $I_{f}$ are assigned a binary value of 1, while those below the threshold are set to 0. The flame boundary is then defined by the contiguous region of pixels with a value of 1. Once the flame boundary is delineated, the flame base and flame tip are tracked to obtain their spatiotemporal response dynamics following the blast wave interaction. Flame height is estimated as the distance between the flame tip and the flame base. The OH* chemiluminescence signal of the flame is determined by integrating the pixel intensities across the obtained flame boundary in the original OH* chemiluminescence image. In the discussions that follow in the subsequent sections, all the flame descriptors presented (flame tip-base spatial positions, flame height, OH* chemiluminescence) are values averaged out over at least three repetitive trials.
    
    The evolution of the blast wave is monitored by spatially tracking its position with time from the recorded Schlieren images. The blast wave Mach number ($M_{s}$) and radius ($R_{s}$) are used as descriptors to characterize the blast wave. The observed blast wave evolution is compared against the analytical blast wave solution formulated by Bach and Lee \cite{bach_analytical_1970} and is plotted in \textbf{Fig. \ref{fig:expt_bach}}. The analytical model assumes that the total mass and energy contained within the blast wave boundary to remain constant during the evolution process, effectively neglecting the effect of air entertainment from the surroundings. Despite this assumption, the observed temporal trends in $M_{s}$ and $R_{s}$ match fairly well with the analytical estimations in the initial phase of the blast propagation (at short time scales; $\sim 1 \, ms$). Due to the experimental limitations of the current facility in acquiring high-speed real-time data on the velocity and pressure fields imposed by the blast wave on the premixed jet flame, the scales obtained from the analytical models are used to correlate and justify the some observed response trends of the premixed jet flame following the interaction process.
    
    }

    \section{Results and Discussion}

    \subsection{Global Observations}\label{subsec:global Obs}
    {
    
    The blast wave generated at the copper wire (that is connected to the high-voltage electrodes) travels radially outward and interacts with the premixed jet flame stabilized at the tip of the central tube. The temporal flow field profiles (velocity and pressure) imposed by the blast wave at the flame base location (tip of the central tube), as estimated from the simplified blast wave model \cite{bach_analytical_1970}, are plotted in \textbf{Fig. \ref{fig:BL_profiles}}. The expanding blast wave reaches the flame base location after a time of $\sim 0.5$ ms ($O(10^{-1})$ ms) from the time of the explosion. The instant of interaction, $t_{i}$, is found to decrease with an increase in the reference blast wave Mach number ($M_{s,r}$) (increase in the charging voltage), as illustrated in \textbf{Fig. \ref{fig:BL_profiles}} and supplementary Figure \textbf{S2}. The imposed velocity ($v_{s}$) and pressure ($p_{s}$) fields exhibit a peak ($v_{p}$ for the velocity field and $p_{p}$ for the pressure field) corresponding to the instant of interaction and are followed by a decaying profile. Both $v_{p}$ and $p_{p}$ are found to increase with $M_{s,r}$. It is interesting to note that these estimated flow fields decay to ambient levels in around $\sim 1$ ms ($\sim O(10^{0})$ ms) from the time instant of explosion. 
    The time instant at which the imposed velocity field decays to ambient levels, $t_{s,v}$, is found to decrease with increasing $M_{s,r}$ (\textbf{Fig. \ref{fig:BL_profiles}}). 
    Similar trends are observed between $t_{s,p}$ (time instant corresponding to the decay of the imposed pressure fields to below ambient levels) and $M_{s,r}$.
    However, the decay time scales associated with the velocity ($t_{s,v}-t_{i}$) and pressure ($t_{s,p}-t_{i}$) fields are found to weakly increase with $M_{s,r}$ (see supplementary Figure \textbf{S2}).
    
    \begin{figure}[!h]
        \centering
        \includegraphics[width=0.9\linewidth]{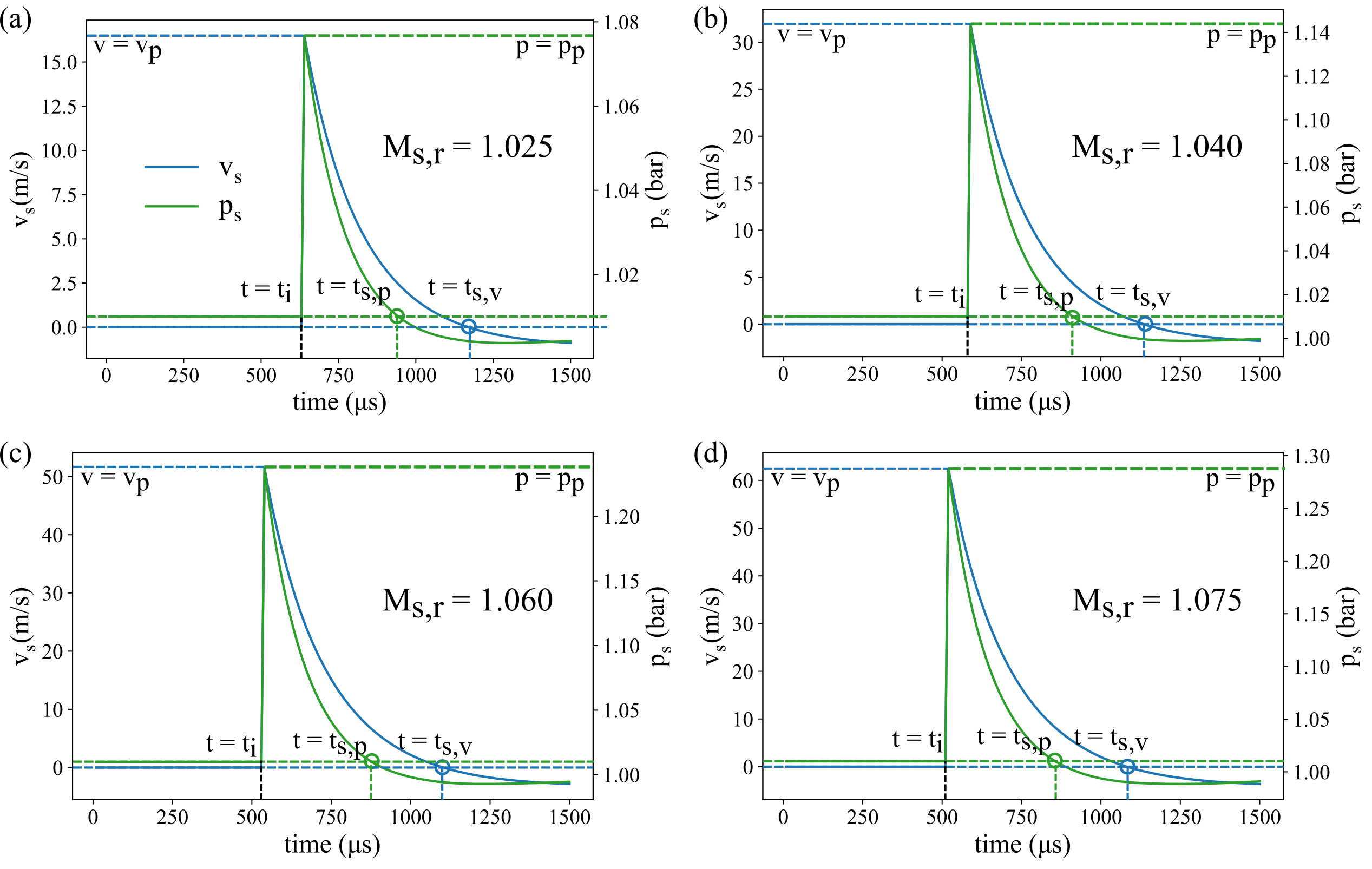}
        \caption{(a-d) Velocity and pressure imposed by the blast wave at the flame base location as estimated using the analytical blast wave model \cite{bach_analytical_1970}. Panels (a-d) correspond to electrode charging voltages of 4kV to 7kV, respectively.}
        \label{fig:BL_profiles}
    \end{figure}
    
    \textbf{Figure \ref{fig:All_Flames} (a)} shows the general schematic of the shock-premixed jet flame interaction process. \textbf{Fig. \ref{fig:All_Flames} (b)} shows the time-resolved response of the flame between $t_i$ and $t_{s,v}$, wherein the blast-imposed velocity ($v_{s}$) exceeds quiescent conditions and induces a characteristic jittery motion of the jet flame. 
    
    \begin{figure*}[!h]
        \centering
        \includegraphics[width=0.7\linewidth]{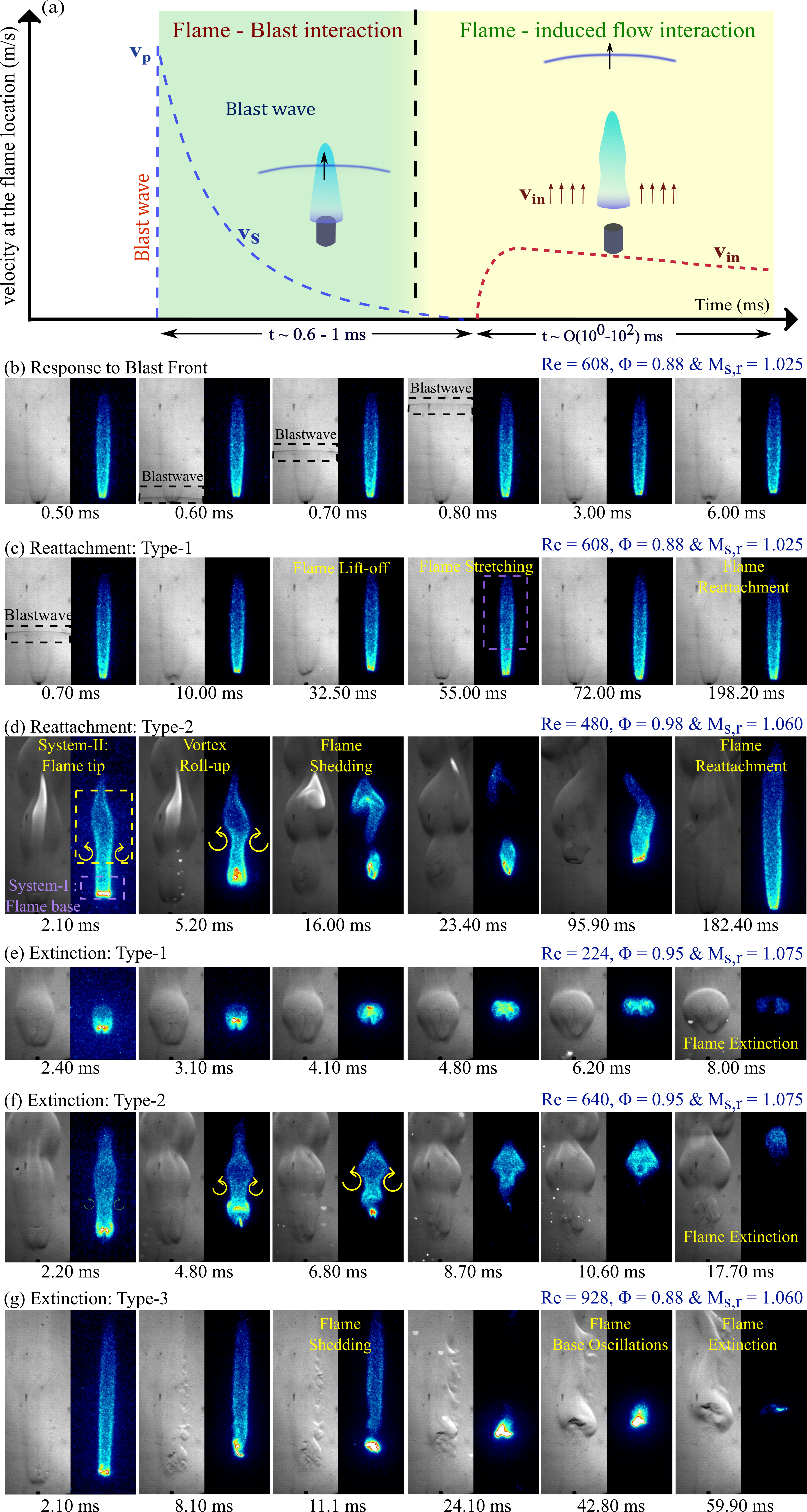}
        \caption{(a) Schematic depicting the temporal variation of the velocity at the flame location. (b-g) Time series of the Schlieren visualization (left) and OH* chemiluminescence (right) images of the flame response to the blast front and the induced flow following it. (b) Flame response to decaying flow field immediately following the blast front. (c,d) Re-attachment Response Regimes: Type-1 and Type-2, respectively. (e-g) Extinction sub-regimes of Type-1, Type-2 and Type-3 respectively. Supplementary Movies (1-5) illustrate the re-attachment and extinction regimes depicted in panels (c-g), respectively.}
        \label{fig:All_Flames}
    \end{figure*}
    
    However, even after the flow field induced by the blast wave decays to quiescent conditions ($t > t_{s,v}$), a pronounced flame response persists, characterized by a distinct flame-base lift-off. This behaviour is found to remain consistent across the entire range of parametric space explored in the current work. This delayed response of the jet flame is hypothesized to be the effect of the bulk flow induced behind the blast wave. As depicted in \textbf{Fig. \ref{fig:BL_profiles}}, the static pressure imposed by the blast wave decays to ambient levels (following an initial spike) in time scales of the order of $\sim 1$ ms from the time of the explosion (denoted as $t_{s,p}$ in the figure). However, this decay continues beyond $t_{s,p}$, eventually dipping below quiescent limits. This sub-quiescent pressure induces flow entertainment from the surroundings. Beyond this limit ($t>t_{s,p}$), the analytical blast wave solution no longer represents the observed flow field since the model neglects the effects of flow entertainment from the surroundings.
    
    Owing to the experimental limitations and shortcomings of the analytical model, accurate estimation of the bulk flow velocity ($v_{in}$) was not possible. However, the scale of the induced bulk flow was estimated based on the initial flame base lift-off rate as illustrated in \textbf{Fig \ref{fig:vin_sch_var} (a)}.
    
    \begin{figure}[!h]
        \centering
        \includegraphics[width=0.75\linewidth]{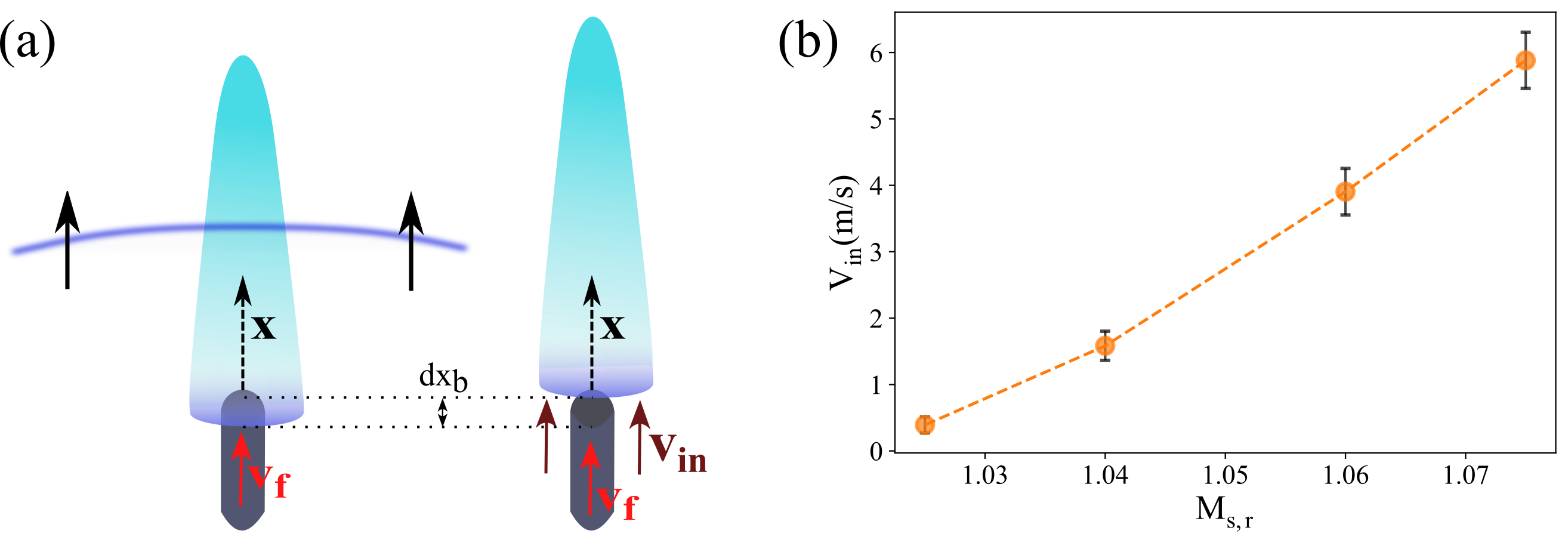}
        \caption{(a) Schematic depicting the flame base lift-off due to the bulk flow induced behind the blast wave. (b) Variation of induced velocity ($v_{in}$) against $M_{s,r}$. The flame base lift-off rate is a near-constant value across the parametric space of fuel-air jet Reynolds number and equivalence ratio for a given strength of the incident blast wave (supplementary Figure \textbf{S3}).}
        \label{fig:vin_sch_var}
    \end{figure}
    
    Prior to the interaction with the blast front and the induced flow following it, the flame was anchored at the nozzle tip, wherein it was stabilized in a narrow mixing layer between the premixed jet and the ambient (\cite{li_stabilization_2022}), at a distance wherein the heat losses balance out the energy production rate at the flame (locally). At the instant where the flame base begins to lift off during its interaction with the induced bulk flow, the local balance at the mixing layer is lost, and the flame boundary tends to advect along with the bulk flow. Hence, its ascent rate $\Bigl(\Big|\frac{dx_{b}}{dt}\Big|_{t_{0}}\Bigr)$, at the instant of lift-off ($t_{0}$), can be scaled with the velocity associated with the induced bulk flow. The method proved effective since the flame base lift-off rate at $t_{0}$ at a given blast strength (given $M_{s,r}$) was found to remain at a near-constant value across the parametric space of $Re$ and $\Phi$ of the premixed jet flame (see supplementary Figure \textbf{S3}). The variation of $v_{in}$ against $M_{s,r}$ is plotted in \textbf{Fig \ref{fig:vin_sch_var} (b)}.
    
    \begin{equation*}
        \Big|\frac{dx_{b}}{dt}\Big|_{t_{0}} \sim v_{in}(x_{0},t_{0})
    \end{equation*}
    
    In the above equation, $x_{0}$ is the location of the flame base at the instant where the flame base begins to lift off at $t_{0}$. $v_{in}$ is found to increase with the blast wave Mach number ($M_{s,r}$) (\textbf{Fig \ref{fig:vin_sch_var} (b)}). The time scale associated with the decay of the induced bulk flow is expected to scale with the flame base lift-off time scales, which is $\sim O(10^2)\,ms$ (almost two orders higher in magnitude in comparison with $t_{s,p}$).
    
    It is to be noted that, as the flame base continues to lift off beyond $t_{0}$, air from the surroundings is continuously entrained into the premixed jet, dynamically altering the global equivalence at the flame base ($\Phi_{b}(x,t)$) and effectively changing the global flame speed ($S_{L,b}(x,t)$) with which the flame base propagates upstream. Additionally, the entertainment also changes the effective velocity of the fuel-air mixture ($v_{f}(x,t)$) directed towards the flame base. The dynamics of flame base lift-off beyond $t_{0}$ are thus a function of $S_{L,b}(x,t)$ and $v_{f}(x,t)$, and a simplified formulation for the same is detailed in section \ref{subsec:Flame base}. 
    
    \textbf{Fig. \ref{fig:All_Flames} (c-f)} presents a time-resolved sequence of OH chemiluminescence and Schlieren flow visualization images, capturing the premixed jet flame's dynamic response to the induced flow following the blast wave. The flame is found to exhibit two distinctive global behaviours: re-attachment (\textbf{Fig. \ref{fig:All_Flames} (c-d)}) and extinction (\textbf{Fig. \ref{fig:All_Flames} (e-f)}), contingent on the operating equivalence ratio ($\Phi$), premixed jet Reynolds number ($Re$), and the incident blast wave Mach number ($M_{s,r}$). While the weakest blast wave ($M_{s,r}$=1.025) caused the flame to re-attach following lift-off, the strongest blast ($M_{s,r}$=1.075) was found to extinguish the flame (across the space of $Re$ and $\Phi$). The flame height (a function of $Re$ and $\Phi$) was also found to play a significant role in the flame response behaviour. Longer flames (higher flame heights) were found to exhibit lower tendencies to undergo extinction for a given incident blast wave Mach number ($M_{s,r}$). An increase in blast wave Mach numbers ($M_{s,r}$) were also found to distort the flame and promote flame pinch-off or flame-shedding events in longer flames. Accordingly, based on the tip response, the re-attachment and extinction regimes were further divided into sub-regimes; Type-1 with negligible flame tip distortion and Type-2 with significant flame tip distortion and necking that might be followed by a flame shedding event. 
    
    Additionally, in a characteristic range of $Re$, $\Phi$ and $M_{s,r}$, the lifted flame was found to exhibit flame base oscillation over prolonged time scales (in comparison with $t_{s,p}$), resulting in a unique extinction pattern (Type-3), typically not observed in droplet flames incident with blast waves (\cite{vadlamudi_insights_2024}). The observed regimes (\textbf{Fig. \ref{fig:All_Flames} (c-f)}) are detailed in section \ref{subsec:Regime Map}
        
    }
    
    \subsection{Regimes of flame response\label{subsec:Regime Map}} \addvspace{10pt}
    
\begin{figure}[!t]
        \centering
        \includegraphics[width=0.9\linewidth]{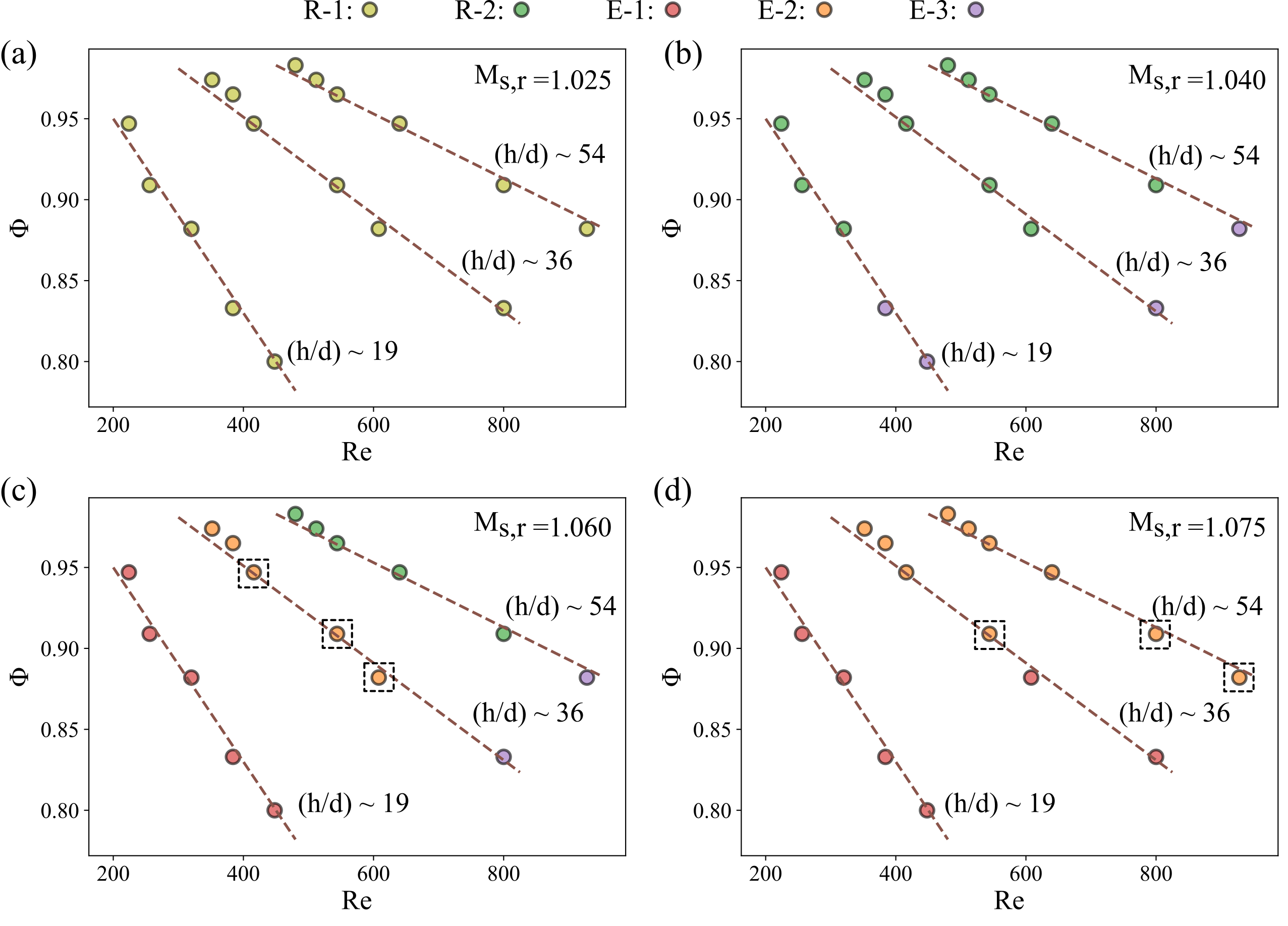}
        \caption{(a-d) Regime map depicting the occurrence of five flame response regimes: R-1, R-2, E-1, E-2, and E-3 in the parametric space of $M_{s,r}$, $Re$ and $\Phi$. The operating points enclosed within dotted rectangles denote conditions where flame tip pinch-off consistently occurred in all experimental trials within Regime E-2. Conversely, other operating points within E-2 exhibited a stochastic pinch-off behaviour.}
        \label{fig:Regime_Map}
    \end{figure}

    \textbf{Fig. \ref{fig:All_Flames} (c)} illustrates the Reattachment: Type-1 (R-1) regime, wherein the flame base lifts off, ascending to a height of $h_{b,lft}$, and exhibits mild stretching at the flame tip in response to the induced flow behind the blast wave. It then re-attaches back at the nozzle tip after a time period of $t_{ra}$. The regime is observed only at the lowest blast strength ($M_{s,r}$=1.025) explored in the current study (\textbf{Fig. \ref{fig:Regime_Map} (a)}). As the strength of the blast wave increases from its lowest value ($M_{s,r} \sim$ 1.04 $\&$ 1.06), the flame base liftoff increases significantly and becomes of the order of the flame height ($h_{b,lft} \sim O(h)$). The flame tip exhibits undulation and violent shedding due to the pronounced circulation buildup in the shear boundary surrounding the flame (explored in detail in Section \ref{subsec:Flame Tip}). This type of flame response regime is classified as Reattachment: Type-2 (R-2) (\textbf{Fig. \ref{fig:All_Flames} (d)}). This response regime prevailed across all flame heights with $\Phi \geq$ 0.88 for incident blast wave Mach numbers of 1.04 and was exhibited by the longest flame (normalized height, $h/d$ $\sim$ 54) when subjected to a blast wave Mach number of 1.06 ($\Phi \geq$ 0.91) (\textbf{Fig. \ref{fig:Regime_Map} (b,c)}).
    
    Jet flames characterized by low flame heights (normalized height, $h/d$ $\sim$ 19), when incident with high strength blast waves ($M_{s,r} \geq$ 1.06), showed continuous flame base lift-off, followed by an extinction event upon flame base-flame tip interaction (\textbf{Fig. \ref{fig:All_Flames} (e)}). The flame tip, Similar to Regime R-1, displayed mild stretching and distortion and was not accompanied by a flame pinch-off event. This response behaviour is termed Extinction: Type-1 (E-1). The regime was also observed at $h/d$ $\sim$ 36 at $\Phi \leq$ 0.83 at blast wave Mach numbers of 1.075. 
    
    In regime E-2 (Extinction: Type-2), which was exhibited by jet flames with higher flame lengths (normalized height, $h/d$ $\sim$ 36 $\&$ 54), there were significant flame tip undulations, resulting in a partial or a complete flame tip pinch-off. Nevertheless, the flame base lift-off in this regime was significant enough ($M_{s,r} \sim 1.075$) to cause the interaction between the flame base and the flame pinch-off point (necking region), which was found to lead to an extinction phenomenon (\textbf{Fig. \ref{fig:All_Flames} (f)}). Another type of flame extinction is observed only at low equivalence ratios ($\Phi$ $\leq$ 0.88) for $1.04 \leq M_{s,r} \leq 1.06$ and is shown in \textbf{Fig. \ref{fig:All_Flames}(g)}. The flame base is observed to lift off and remain at a near-constant liftoff height for prolonged periods of time (compared to E-2) before undergoing extinction. Similar to that observed in Regime E-2, a flame pinch-off event is observed.  However, a key distinction emerges following pinch-off. Unlike E-2, following flame pinch-off, an intense flame base is observed, which exhibits significant oscillations in intensity and shape. This flame response regime is classified as Extinction: Type-3 (E-3).
    
    \begin{figure*}
        \centering
        \includegraphics[width=1\linewidth]{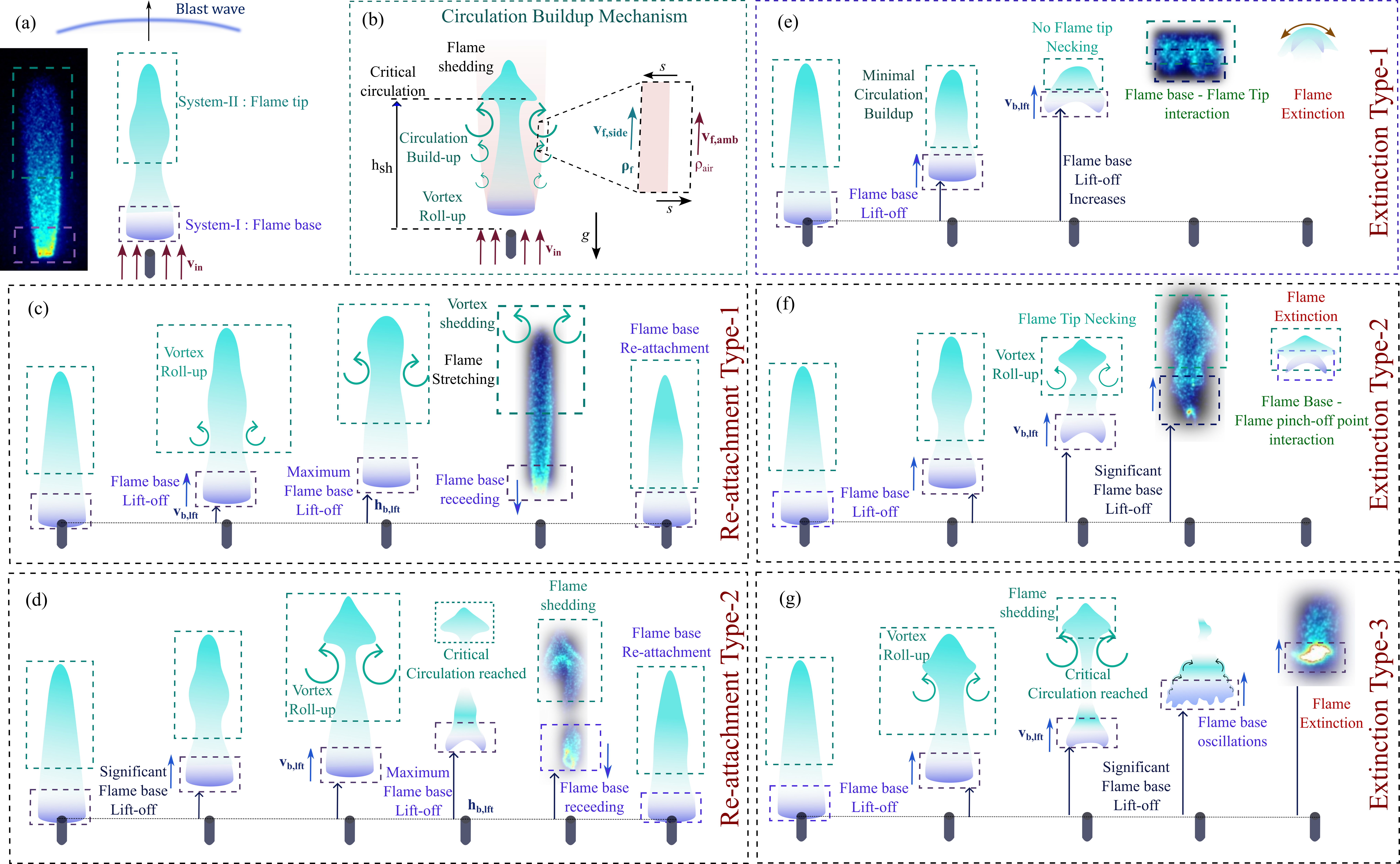}
        \caption{(a) Schematic depicting system-I (flame base) and system-II (flame tip) of the premixed jet flame. (b) Schematic illustrating the effect of induced flow on flame. (c,d) A sequence of schematics depicting the two reattachment regimes: R-1 and R-2, respectively. (e-g) A sequence of schematics depicting the three extinction regimes: E-1, E-2 and E-3, respectively}
        \label{fig:Schematic}
    \end{figure*}
    
\begin{figure*}
        \centering
        \includegraphics[width=1\linewidth]{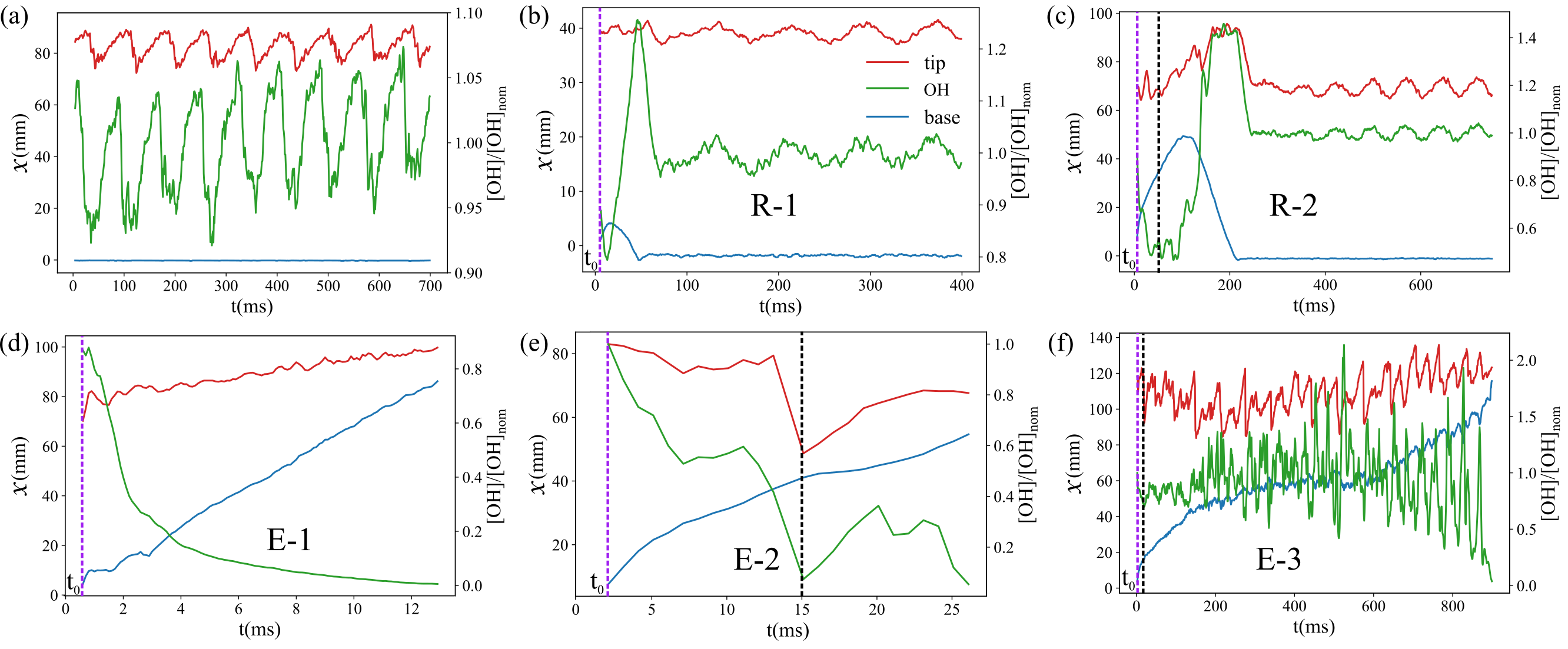}
        \caption{The temporal response of the flame tip (red) and the flame base (blue) is depicted alongside the flame's OH* chemiluminescence signal (normalized with the average OH* chemiluminescence intensity of the unforced flame at the same $Re$ and $\Phi$) under quiescent conditions and different flame response regimes. (a) Buoyant flickering of the jet flame in a quiescent environment at $Re = 544$ and $\Phi = 0.96$. (b) R-1 Regime at $Re = 256$, $\Phi = 0.91$ and $M_{s,r} = 1.025$. (c) (b) R-2 Regime at $Re = 416$, $\Phi = 0.95$ and $M_{s,r} = 1.040$. (d) E-1 Regime at $Re = 800$, $\Phi = 0.83$ and $M_{s,r} = 1.075$. (e) E-2 Regime at $Re = 544$, $\Phi = 0.91$ and $M_{s,r} = 1.060$. (f) E-3 Regime at $Re = 928$, $\Phi = 0.88$ and $M_{s,r} = 1.040$. Purple dotted lines in the figures correspond to the time instant where the flame base starts to lift off ($t_{0}$), and dotted black lines correspond to the time instant where flame shedding is observed.}
        \label{fig:quant_response}
    \end{figure*}

    To better understand the dynamics of the interaction process, the flame base (system-I) and tip (system-II) responses to the induced flow are studied separately. \textbf{Fig. \ref{fig:Schematic}} shows a schematic depicting the observed flame response regimes, breaking them down into flame base response and flame tip response. A quantitative representation of the same is provided in \textbf{Fig. \ref{fig:quant_response}}, wherein the temporal response of the flame base and flame tip is plotted separately alongside the OH* chemiluminescence signal of the flame. 
    While the flame base is observed to lift off across all regimes, it only re-attaches back at the nozzle tip in R-1 and R-2 (\textbf{Fig. \ref{fig:Schematic}(c,d)}, plotted in blue). In these regimes (\textbf{Fig. \ref{fig:quant_response}(b,c)}), the flame base lifts off to reach a height of $h_{b,lft}$ and then re-attaches back at the nozzle tip after a time of $t_{ra}$ ($t_{0}$ is the reference from which $t_{ra}$ is measured). A simplified formulation to establish the dependencies of these characteristic parameters on $Re$, $M_{s,r}$ and $\Phi$ is provided in Section \ref{subsec:Flame base}. 
    In R-1, the flame tip displays slight undulations before returning to its characteristic buoyant flickering behaviour, whereas in R-2, significant flame distortion occurs, accompanied by a flame pinch-off event, before reverting to its buoyant flickering state (\textbf{Fig. \ref{fig:quant_response}(b,c)}, plotted in orange). 
    Buoyant flickering arises as a consequence of buoyancy-induced instability, which leads to the rolling up of vortices along the shear boundary (between the hot product gases and the ambient air) surrounding the flame. As these vortices convect downstream along the shear boundary, they grow and eventually shed when their circulation reaches a critical limit. This shedding process, occurring periodically, is accompanied by distortion of the flame tip, resulting in periodic flame flickering (\textbf{Fig. \ref{fig:Schematic}(b)}, \textbf{Fig. \ref{fig:quant_response}(a)},).
    The bulk flow following the blast front imposes a velocity differential across the shear boundary and adds to the circulation build-up rate of the shear layer vortices. This causes them to reach their critical circulation limit in shorter timescales, increasing the flickering frequency, $f_{sh}$, and reducing the advective length scale (flame shedding height, $h_{sh}$) at which shedding is observed. A simplified formulation illustrating the dependence of $h_{sh}$ and $f_{sh}$ on $v_{in}$ is provided in Section \ref{subsec:Flame Tip}. 
    As the R-2 regime is typically observed at higher shock strengths (higher $v_{in}$), the associated flame tip distortion, increase in flickering frequency (decrease in flickering time scales), and decrease in shedding heights become significantly more pronounced compared to those observed in the R-1 regime (\textbf{Fig. \ref{fig:quant_response}(b,c)}, plotted in orange). Notably, in R-2, critical circulation limits are attained at heights considerably lower than the nominal flame heights, leading to the formation of a neck and an eventual flame pinch-off event (\textbf{Fig. \ref{fig:Schematic}(d)}). Conversely, in the R-1 regime, $h_{sh}$ is comparable to nominal shedding heights, and the flame tip experiences only a mild stretch following the interaction with the induced flow (\textbf{Fig. \ref{fig:Schematic}(c)}).
    The OH* chemiluminescence signal of the flame, which is found to scale with the instantaneous flame height, is found to attain its global minima at the instant corresponding to the first vortex shedding event following the interaction with the induced flow. It is to be noted that this vortex-shedding process is not accompanied by a flame pinch-off event in R-1, unlike in R-2. The OH* chemiluminescence signal of the flame reverts back to its nominal flickering pattern after the effect of the induced flow has subsided (\textbf{Fig. \ref{fig:quant_response}(b,c)}, plotted in green). An intriguing observation is that there is a significant rise in the OH* chemiluminescence signal following the first vortex-shedding event after the interaction between the jet flame and the induced flow. This enhancement likely corresponds to the restoration of the jet flame height to a level dictated by the local Reynolds number ($Re$) and equivalence ratio ($\Phi$) after the initial disruption caused by the interaction process.
    
    The extinction regimes (E-1, E-2, and E-3) plotted in \textbf{Fig. \ref{fig:quant_response}(d,e,f)} show that the flame base continues to lift off following the interaction with the induced flow without attaining a maxima. In Regime E-1, this leads to the interaction between the flame base and the flame tip, while in Regime E-2, it results in the interaction between the flame base and the flame neck. This interaction has been found to result in flame extinction. We hypothesise that, during the interaction, the advective velocity imposed by the flame base on the flame tip/flame neck causes the flame strain rates to exceed extinction strain rate limits, and this results in flame extinction (\textbf{Fig. \ref{fig:Schematic}(e,f)}).
    It is to be noted that, akin to regime R-2, neck formation (owing to faster circulation buildup, resulting in lower shedding heights) is observed in regime E-2.  
    However, flame pinch-off is not observed across the entire parametric space of E-2. This is due to the presence of an additional time scale, $t_{b,ext}$, that accounts for the time taken by the flame base to advance downstream and interact with the flame neck. If the time required for the circulation build-up to reach its critical limit ($t_{circ} = 1/f_{sh}$) exceeds $t_{b,ext}$, then flame pinch-off is not observed following the neck formation. However, if $t_{circ}<t_{b,ext}$, a flame pinch-off event similar to that observed in the R-2 regime occurs (\textbf{Fig. \ref{fig:quant_response}(e)}). 
    $t_{b,ext}$ is expected to scale directly with the flame height and inversely with the base lift-off velocity, both of which are dependent on $Re$, $M_{s,r}$, and $\Phi$. Similarly, although the dependence of $t_{circ}$ with $M_{s,r}$ is intuitively explained (above) based on the enhanced circulation build-up rate due to induced bulk flow, the inherent circulation build-up mechanism: buoyancy-induced circulation build-up rate is dependent on $Re$ and $\Phi$ (explored in detail in Section \ref{subsec:Flame Tip}). Thus, both competing time scales are functions of $Re$, $\Phi$ and $M_{s,r}$. Regime E-2 is typically observed when high-strength blast waves (wherein high induced velocities inversely affect both $t_{b,ext}$ and $t_{circ}$) are incident on long ($h/d \geq 36$ at $M_{s,r}=1.075$) to intermediate ($h/d \sim 36$ at $M_{s,r}=1.06$) flames. 
    The observed parametric space of E-2 is anticipated to make $t_{circ}$ comparable and of the same order as $t_{b,ext}$, since partitioning the parameter space of E-2 into regimes where pinch-off occurs and where it does was not straightforward and was even found to be stochastic in a certain parametric space within E-2. The operating points enclosed within a dotted rectangle in \textbf{Fig. \ref{fig:Regime_Map}(c,d)} represents operating conditions for which flame pinch-off was consistently observed across all experimental trials, while other operating conditions in E-2 displayed a stochastic pinch-off event across the experimental trials.
    
    Regime E-1 is typically encountered at high shock strengths ($M_{s,r}$) and low flame heights. The high advective velocity of the flame base (at high $M_{s,r}$), alongside low associated flame heights, makes $t_{b,ext}$ significantly lower than $t_{circ}$. This results in flame extinction well before the circulation buildup is sufficient to induce neck formation (\textbf{Fig. \ref{fig:Schematic}(e)}). The plot in \textbf{Fig. \ref{fig:quant_response}(d)} depicts that the OH* chemiluminescence signal of the flame response decays to zero in timescales of $t_{b,ext}$, before shear layer circulation build-up could reach critical limits and result in flame pinch-off event which is usually associated with a drastic distortion in OH* chemiluminescence signal in the flame response phase. 
    
\begin{figure*}
        \centering
        \includegraphics[width=0.9\linewidth]{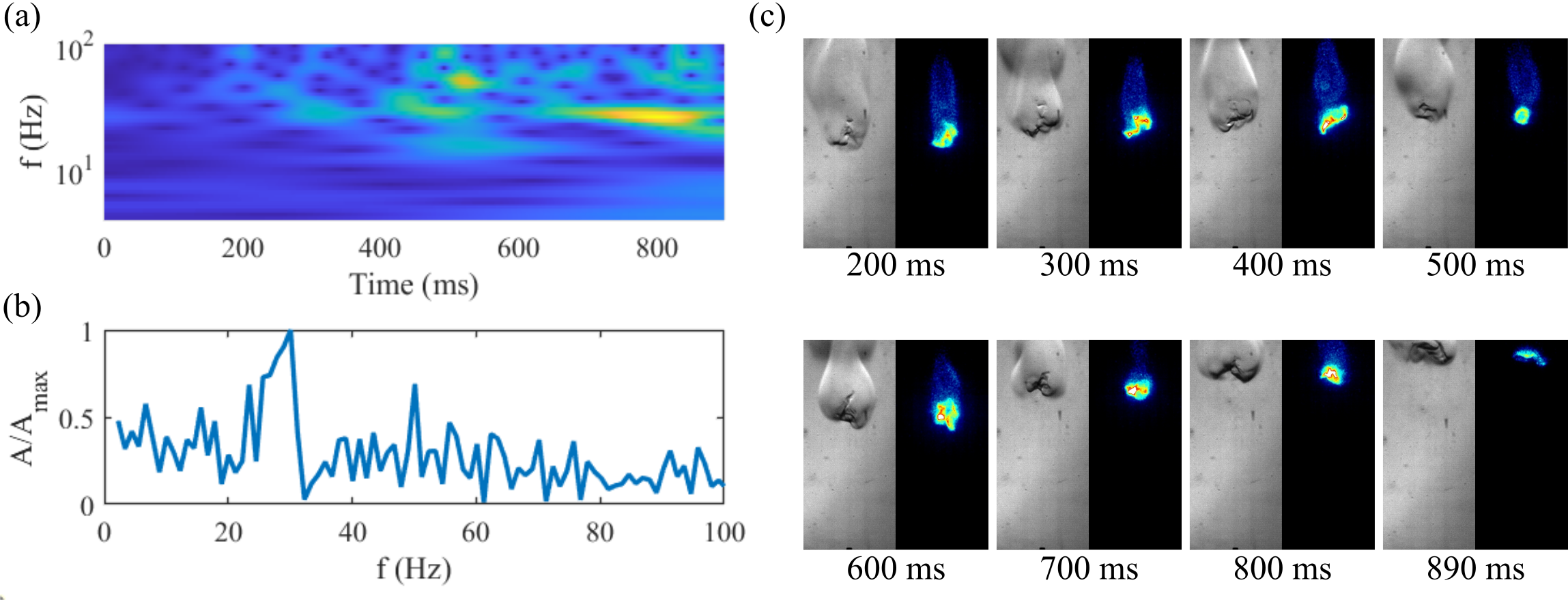}
        \caption{(a,b) Continuous wavelet transform and Fast Fourier transform of the OH* chemiluminescence signal of the flame in the E-3 Response Regime, respectively. The associated time series of the OH* chemiluminescence signal is shown in \textbf{Fig. \ref{fig:quant_response}(f)}, and corresponds to $Re = 928$, $\Phi = 0.88$ and $M_{s,r} = 1.040$. (c) Image sequence depicting the flame base oscillations in the E-3 Regime for the same operating conditions.}
        \label{fig:E3}
    \end{figure*}

    However, unlike the other extinction regimes, E-3 does not exhibit any flame base - flame tip/neck interaction and extinguishes as the flame base continues to lift off gradually, exhibiting intense oscillations in intensity and shape as depicted in \textbf{Fig. \ref{fig:E3}(c)}. However, it is to be noted that, similar to the regime E-2, E-3 exhibits a flame pinch-off event, following which the flame base oscillations are observed (\textbf{Fig. \ref{fig:All_Flames}(g)}). \textbf{Fig. \ref{fig:quant_response}(f)} depicts the time series of the OH* chemiluminescence signal of the flame in the E-3 Regime. A continuous wavelet transform and FFT of the signal reveals that these oscillations are of the order of the flame's flickering frequency ($f_{sh}$) (\textbf{Fig. \ref{fig:E3}(b,c)}). However, these oscillation frequencies were not found to exhibit any discernible trend with $Re$, $\Phi$ and $M_{s,r}$ within the observed parametric space of E-3, and requires further investigation over a wider parametric space at low equivalence ratios. 
    
    \subsection{Flame Base Lift-off\label{subsec:Flame base}} 
\begin{figure*}
        \centering
        \includegraphics[width=0.75\linewidth]{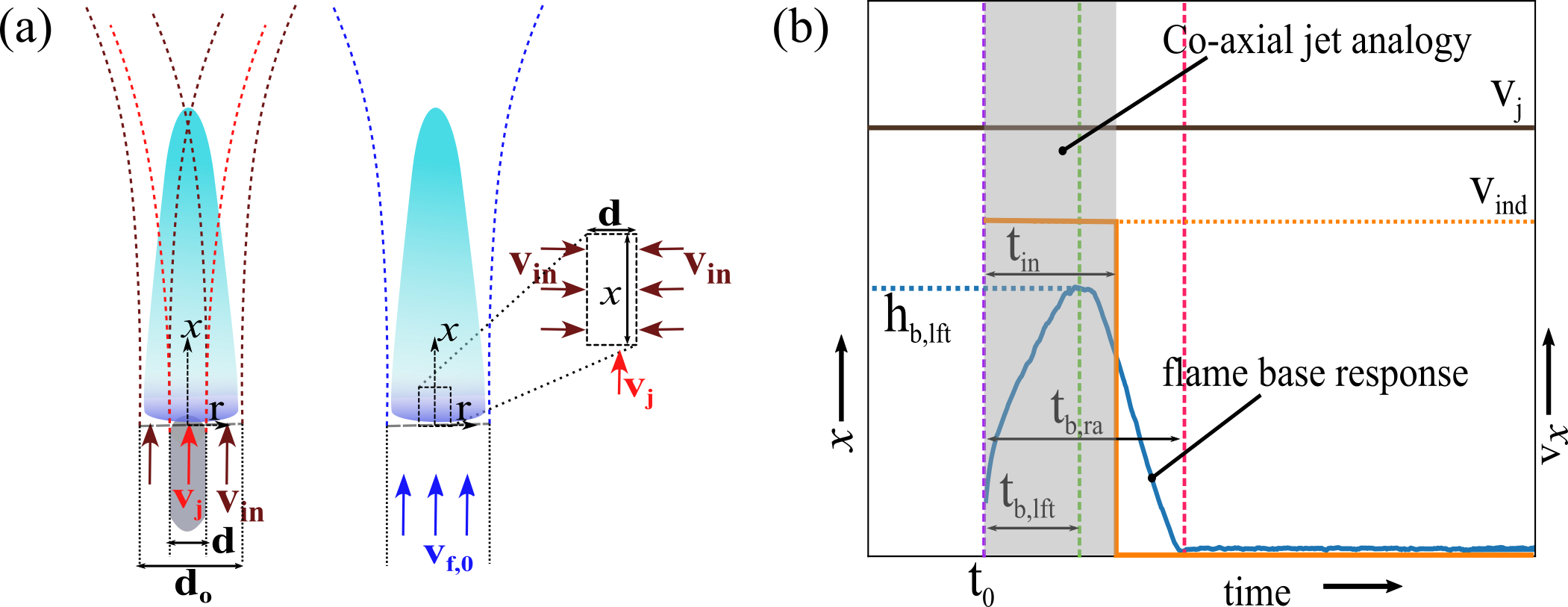}
        \caption{(a) Schematic depicting the co-axial jet analogy of the blast wave - premixed jet flame interaction problem. (b) Sample plot depicting the flame base lift-off behaviour and the characteristics of the co-axial jet approximation.}
        \label{fig:Flame_base_sch}
    \end{figure*}

\addvspace{10pt}
    This section details a simplified model to explain the observed dependencies of the flame base lift-off height ($h_{b,lft}$) and its associated time scale ($t_{b,lft}$) on $Re$, $\Phi$ and $M_{s,r}$. 
    As the induced bulk flow (behind the blast wave) passes across the premixed jet flame, the flame base lifts off. Given the radial-outward propagation geometry of the blast wave, it is anticipated that the induced flow will also propagate radially outward. However, it is expected that only a portion of the induced flow in close proximity to the central jet tube will directly interact with the premixed flame. Additionally, the effects of the induced flow over the dynamics of the premixed flame are expected to subdue in $O(10^{1})$ ms (estimated based on the time scales associated with flame base reattachment in the reattachment regimes).
    Consequently, the problem can be approximated as the response of a premixed flame to an equivalent co-axial impulsive air jet surrounding the central tube, as illustrated in Fig. \ref{fig:Flame_base_sch} (a). To further simplify the formulation, the co-axial air jet, representing the induced flow, is assumed to possess a steady velocity profile for a duration of $t_{in}$ ($O(10^{1})$ ms). Following this time period, the jet velocity is assumed to instantly drop to zero (\textbf{Fig. \ref{fig:Flame_base_sch}(b)}). Thus, for a period between $t_{0}$ (instant of interaction between the induced flow with the premixed flame base) and $(t_{0} + t_{in})$, the problem reduces to a steady-coaxial jet formulation. The inner tube carries a premixed fuel-air jet at an equivalence ratio of $\Phi$ and a mixture velocity of $v_{j}$, while the outer co-axial air jet surrounding the inner tube has a velocity of $v_{in}$. The outer diameter of the co-axial tube ($d_{o}$) characterises the equivalent radial diameter of the induced flow field that interacts with the flame (\textbf{Fig. \ref{fig:Flame_base_sch}(a)}). $d_{o}$ is expected to be of the order of the flame base diameter ($d_{f,b}$).
    
    Following the formulation of \cite{alexander_schumaker_mixing_2012}, a co-axial steady jet can then be approximated into an equivalent single jet having the same momentum flux as that of the co-axial jets.
    
    \begin{gather*}
        \rho_{j} \frac{\pi d^{2}}{4} {v_{j}}^{2} + \rho_{a} \frac{\pi ( {d_{o}^{2}} - {d^{2}} )}{4} {v_{in}}^{2} = \rho_{a} \frac{\pi {d_{o}^{2}}}{4} {v_{f,0}}^{2} \notag \\
        \Rightarrow v_{f,0} = \sqrt{ \Bigl(\frac{\rho_{j}}{\rho_{a}}\Bigr) \Bigl(\frac{d}{d_{o}}\Bigr)^2 v_{j}^2 + \Bigl(\frac{d_{o}^2 - d^2}{d_{o}^2}\Bigr) v_{in}^2 } \tag{1}
    \end{gather*}
    In the above equation, $\rho_{j}$ and $\rho_{a}$ are the densities of the premixed fuel-air mixture in the central tube and the ambient air in the outer co-axial tube, respectively. $v_{f,0}$ corresponds to the velocity of the equivalent single jet at the tube exit.
    
    Thus, the problem at hand has been reduced to an equivalent single jet that accounts for the effects of the induced flow with an equivalent premixture velocity ($v_{f,0}$). However, it is important to note that the equivalent single jet is an open jet that expands radially outward into its surroundings. Consequently, the effective jet velocity becomes a function of both the radial distance from the centre-line ($r$) and the axial distance from the nozzle exit ($x$). This single jet can then be modelled as an expanding Schlichting jet \cite{schlichting_laminare_1933}. The velocity profile of the expanding jet can be formulated as,
    
    \begin{gather*}
        v_{f} (r,x) = \frac{256 \nu}{3} \frac{Re_{f}}{ \Bigl(\frac{32}{3}\Bigr)^2 \Bigl(\frac{x}{Re_{f}}\Bigr) + r\Bigl(\frac{r Re_{f}}{x} \Bigr)^3 + \frac{64r}{3}\Bigl(\frac{r Re_{f}}{x} \Bigr)}
    \end{gather*}
    In the equation, $Re_{f}$ is the equivalent Reynolds number of the single jet (based on $v_{f,0}$). The above formulation can now be used to evaluate the local velocity of the fuel-air mixture at the flame base location. Specifically, this analysis will be used to evaluate the local velocity at a radial distance of $r_{f,b}$, corresponding to the flame base radius ($r_{f,b} = d_{f,b}/2 \sim d_{o}/2$), and an axial distance of $h_{b,lft}$ (denoted as $v_{f}(h_{b,lft},r_{f,b})$). 
    In the re-attachment regimes, when $t = t_{b,lft}$, $x_{b}$ (flame base location) attain a value of $h_{b,lft}$ and $(dx_{b}/dt)_{t_{b,lft}}$ becomes $0$, since the flame base attains its maxima (\textbf{Fig. \ref{fig:Flame_base_sch}(b)}). The observed values of $h_{b,lft}/d$ and $Re_{f}$ are much higher than $1$, wherein $h_{b,lft}/d \sim 0(10^1)$ and $Re_{f} \sim 0(10^2)$. Thus, evaluating $v_{f}(h_{b,lft},r_{f,b})$ in the limit where the flame base reaches its maxima (maximum lift-off height), the following scaling arguments can be derived between the terms of the above equation.
    \begin{gather*}
        r\Bigl(\frac{r Re_{f}}{x} \Bigr)^3 >> \frac{64r}{3}\Bigl(\frac{r Re_{f}}{x} \Bigr) \\
        r\Bigl(\frac{r Re_{f}}{x} \Bigr)^3 >> \Bigl(\frac{32}{3}\Bigr)^2 \Bigl(\frac{x}{Re_{f}}\Bigr)
    \end{gather*}
    $v_{f}(h_{b,lft},r_{f,b})$ can then be simplified as,
    
    \begin{gather*}
        v_{f}(h_{b,lft},r_{f,b}) \sim \frac{(\nu h_{b,lft})^3}{v_{f,0}^2 d^2 d_{o}^4} \tag{2}
    \end{gather*}
    
    Similar to the estimation performed above on the velocity scale, a scalar transport mixing model, as proposed by \cite{villermaux_mixing_2000}, can be used to estimate the equivalent equivalence ratio of the analogous single jet. The model assumes that the outer co-axial jet is entrained into the inner jet at a velocity scale proportional to its axial velocity, $v_{in}$. Thus, for a control volume of height $x$, depicted in \textbf{Fig. \ref{fig:Flame_base_sch}(a)}, the mass flux of the outer jet entrained within the inner jet can be formulated as $\rho_{a} (\pi d) x v_{in}$. The resultant equivalence ratio ($\phi_{b,m}$) following this entertainment can be estimated as,
    \begin{gather*}
        \phi_{b,m}(x) = \frac {\phi} {1 + \Bigl(\frac{\rho_{a}}{\rho_{j}}\Bigr) \Bigl(\frac{v_{in}}{v_{j}}\Bigr) \Bigl(\frac{4x}{d}\Bigr)} \tag{3}
    \end{gather*}
    Although the above formulation accounts for the axial variation of the equivalence ratio, it should be noted that the equivalence ratio also varies along the radial direction. Thus, $\phi_{b,m}(x)$ estimated in the above equation represents the average value of the equivalence ratio at an axial distance of $x$ from the nozzle exit. Following the formulation of \cite{vadlamudi_insights_2023} for co-axial fuel-air jets, a Gaussian profile of equivalence ratio distribution is assumed across the radial direction, which yields the following expression for equivalence ratio as a function of $r$ and $x$:
    \begin{gather*}
        \phi_{b}(r,x) = \phi \, exp \Bigl(\frac{-(\phi (r/r_{s}))^2}{18 \phi_{b,m}^2}\Bigr)
    \end{gather*}
    In the above equation, $r_{s}$ is the length scale that quantifies the radial spread of the assumed Gaussian profile for the equivalence ratio. An estimate of $r_{s}$ can be obtained based on the length scale associated with mass diffusion during the period of flame base lift-off (timescale of $t_{b,lft}$). 
    \begin{gather*}
        r_{s} \sim (D t_{b,lft})^{1/2}
    \end{gather*}
    A scale for $D$ can be obtained based on the binary diffusion coefficient of methane in air, which is of the order of $O(10^{-5}) \, m^2/s$ \cite{langenberg_technical_2020}. Experimental observations show that $t_{b,lft} \sim O(10^{-1}) \, s$. Thus, $r_{s}$ is expected to scale as $0(10^{-3}) \, m$. 
    Substituting for $\phi_{b,m} (x)$ at $x = h_{b,lft}$ into the above equation in the limit where $(h_{b,lft}/d) \sim O(10^{1})$, we get,
    \begin{gather*}
        \phi_{b}(h_{b,lft},r_{f,b}) \sim \phi \, exp \Bigl( -\Bigl[ \Bigl(\frac{d_{o}}{2 r_{s}}\Bigr) \Bigl(\frac{\rho_a}{\rho_j}\Bigr) \Bigl(\frac{v_{in}}{v_{j}}\Bigr) \Bigl(\frac{h_{b,lft}}{d}\Bigr) \Bigr]^2 \Bigr) \tag{4}
    \end{gather*} \label{eq_phi_eq}
    The obtained estimate of the equivalence ratio can now be used to estimate the flame speed at the flame base using flame speed co-relations for methane-air flames, as proposed by Gulder et al \cite{gulder_correlations_1984}.
    \begin{gather*}
        S_{L} = 0.422 \phi^{0.15} \, exp (-5.18 (\phi - 1.075)^2 ) \, m/s
    \end{gather*}
    It should also be noted that lifted flames are stabilised with an edge flame at its base (\cite{buckmaster_edge-flames_2002,vadlamudi_insights_2023}) whose effective flame speed ($S_{L,b}$) is a multiple (say, $A$) of the laminar flame speed ($S_{L}$) at that equivalence ratio (\cite{buckmaster_edge-flames_2002}). A formulation for $S_{L,b}$ at $x=h_{b,lft}$ can thus be obtained by substituting for $\phi$ from \textbf{Eq. (4)} into the above equation. 
    \begin{gather*}
        S_{L,b} (h_{b,lft},r_{f,b}) = A \, (0.422 \phi_{b}(h_{b,lft},r_{f,b})^{0.15} \, exp (-5.18 (\phi_{b}(h_{b,lft},r_{f,b}) - 1.075)^2)) \tag{5}
    \end{gather*}
    
\begin{figure*}
        \centering
        \includegraphics[width=0.9\linewidth]{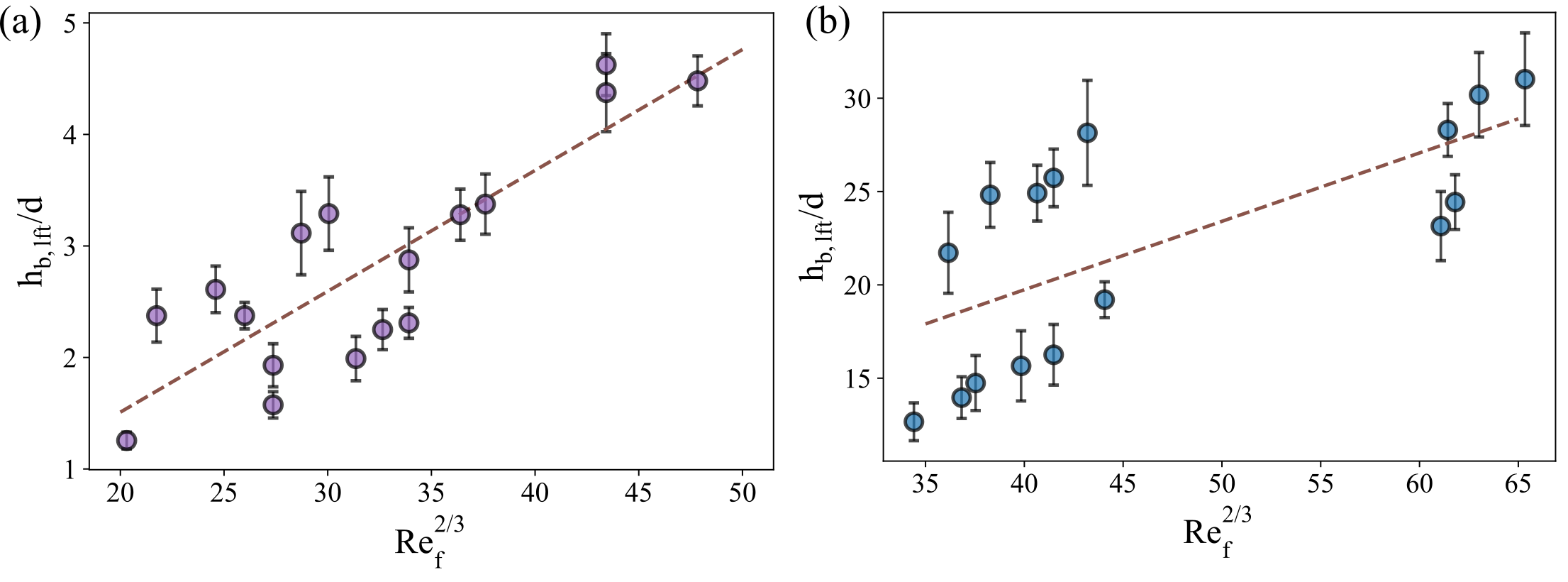}
        \caption{(a,b) $(h_{b,lft}/d)$ is plotted against $Re_{f}^{2/3}$, where, $Re_{f} = v_{f,0}(d/\nu)$, in the R-1 and R-2 reattachment regimes, respectively.}
        \label{fig:flame_hbmax}
    \end{figure*}

    Since the lifted flame reaches its maxima at $t = t_{b,lft}$, wherein $(dx_{b}/dt)_{t_{b,lft}} = 0$, the flame attains a state of quasi-equilibrium at that instant, wherein the local flow speed ($v_{f}(h_{b,lft},r_{f,b})$) balances out the local flame propagation velocity ($S_{L,b} (h_{b,lft},r_{f,b})$).
    \begin{gather*}
        \Big| S_{L,b} (h_{b,lft},r_{f,b}) \sim v_{f}(h_{b,lft},r_{f,b}) \Big|_{t_{b,lft}}
    \end{gather*}
    Simplifying the above equation yields the following correlation for $h_{b,lft}$
    \begin{gather*}
        h_{b,lft}^3 \sim \frac{v_{f,0}^2 d^2 d_{0}^4}{\nu^3} \, A (0.422 \phi_{b}(h_{b,lft},r_{f,b})^{0.15} \, exp (-5.18 (\phi_{b}(h_{b,lft},r_{f,b}) - 1.075)^2))
    \end{gather*}
    It should be noted that the term, $\Bigl[\Bigl(\frac{d_{o}}{2 r_{s}}\Bigr) \Bigl(\frac{\rho_a}{\rho_j}\Bigr) \Bigl(\frac{v_{in}}{v_{j}}\Bigr) \Bigl(\frac{h_{b,lft}}{d}\Bigr) \Bigr]$ exceeds $1$, given that $(h_{b,lft}/d)$ is of the order of $0(10^{1})$ and other ratios in the term lie between $0(10^{-1})$ and $0(10^{0})$.
    This causes $A (0.422 \phi_{b}(h_{b,lft},r_{f,b})^{0.15} \, exp (-5.18 (\phi_{b}(h_{b,lft},r_{f,b}) - 1.075)^2))$ to tend to a small positive constant value (say, $k$) of the order of $0(10^{-2})$ across the entire parametric space.
    This reduces the above scaling law to,
    \begin{gather*}
        h_{b,lft}^3 \sim k \frac{v_{f,0}^2 d^2 d_{0}^4}{\nu^3} \tag{5}
    \end{gather*}
    Thus, $h_{b,lft}^3$ is expected to scale directly with $v_{f,0}^2$ and is evident in \textbf{Fig. \ref{fig:flame_hbmax}(a,b)}. 
    
    The plots suggest a near-linear correlation between $h_{b,lft}$ and $v_{f,0}^{2/3}$, which can be expressed as $h_{b,lft} = C(v_{f,0}^{2/3}) + D$ (or $h_{b,lft}/d = E(Re_{f,0}^{2/3}) + F$), where $C$ and $D$ are fitting constants. Linear fit curves for each of the reattachment regimes (R-1 and R-2) are plotted in \textbf{Fig. \ref{fig:flame_hbmax}(a,b)}, respectively. 

    Following similar lines, we can also obtain the scale for the time associated with flame base lift-off ($t_{b,lft}$), which can be formulated as follows:
    \begin{gather*}
        t_{b,lft} \sim \frac{h_{b,lft}}{v_{f}(h_{b,lft},r_{f,b})}
    \end{gather*}
    Substituting for $v_{f}(h_{b,lft},r_{f,b})$ in terms of $h_{b,lft}$, we get,
    \begin{gather*}
        t_{b,lft} \sim \frac{v_{f,0}^2 d^2 d_{o}^4}{\nu^3 h_{b,lft}^2}
    \end{gather*}
    Invoking the dependence of $h_{b,lft}$ on $v_{f,0}$ (\textbf{Eq. (5)}), we can obtain the following scaling law.
    \begin{gather*}
        t_{b,lft} \sim \frac{d^{2/3} d_{o}^{4/3}}{\nu k^{2/3}} v_{f,0}^{2/3} \tag{6}
    \end{gather*}
 
\begin{figure*}
        \centering
        \includegraphics[width=0.9\linewidth]{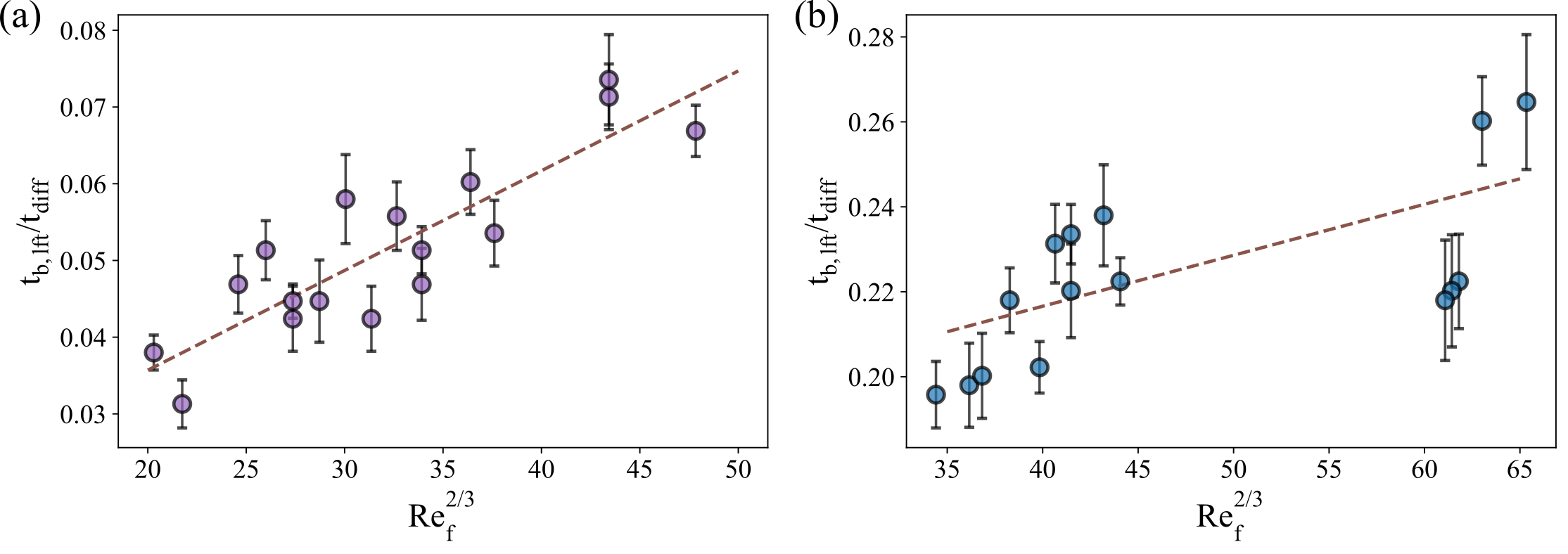}
        \caption{(a,b) $(t_{ra}/t_{diff})$ is plotted against $Re_{f}^{2/3}$, where, $Re_{f} = v_{f,0}(d/\nu)$, in the R-1 and R-2 reattachment regimes, respectively.}
        \label{fig:flame_tblft}
    \end{figure*}

    Experimental plots in \textbf{Fig. \ref{fig:flame_tblft} (a,b)} also illustrate the above dependence. The plots depict a linear fit between $(t_{b,lft}/t_{diff})$ and $Re_{f}^{2/3}$, similar to that observed between $(h_{b,lft}/d)$ and $Re_{f}^{2/3}$ in \textbf{Fig. \ref{fig:flame_hbmax}}. It is to be noted that, in the plots, $t_{diff}$ is the characteristic diffusion time scale and is estimated as $t_{diff} = d^2/D$. $t_{diff}$ remains constants across the explored parametric space.  

\begin{figure*}
        \centering
        \includegraphics[width=0.9\linewidth]{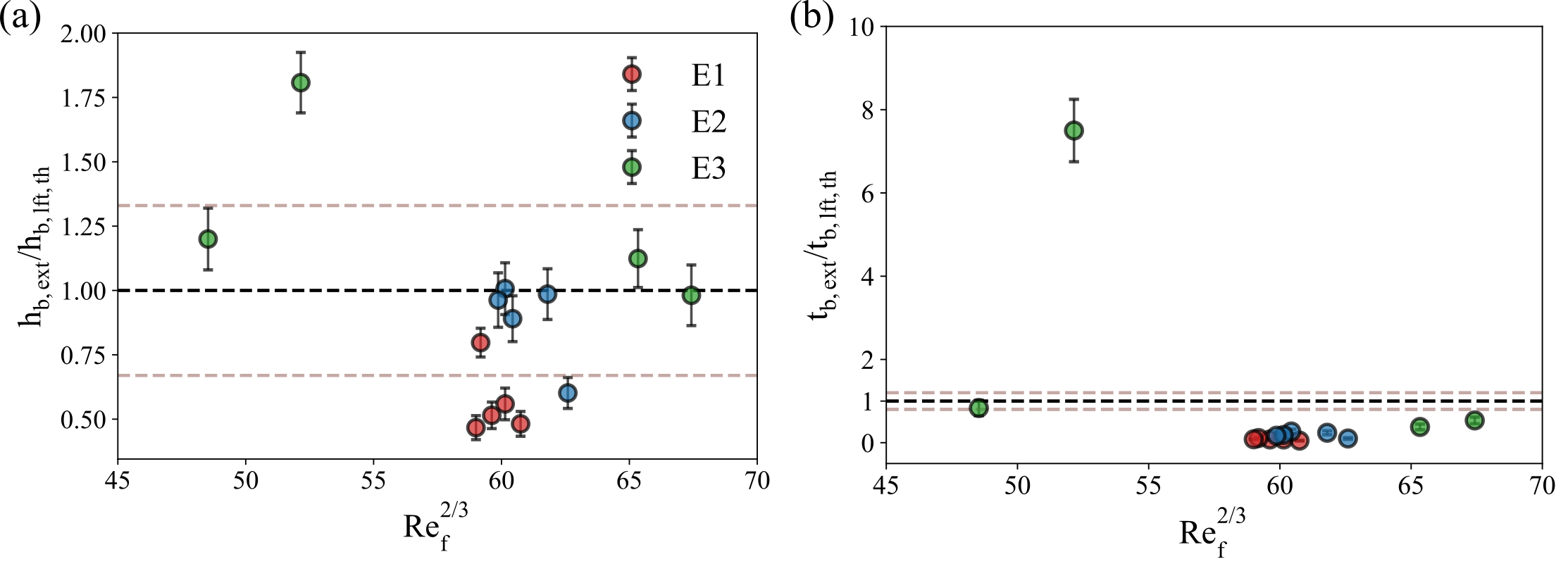}
        \caption{(a,b) $h_{b,ext}/h_{b,lft,th}$ and $t_{b,ext}/t_{b,lft,th}$ are plotted against $Re_{f}^{2/3}$, respectively. Both the plots correspond to the extinction regimes}
        \label{fig:scale_hblft_tra_ext}
    \end{figure*}

    We can now extend the above formulation and the experimentally observed correlations between $(h_{b,lft}/d)$ and $(t_{b,lft}/t_{diff})$ against $Re_{f}^{2/3}$ to the extinction regimes and compare these theoretical scales with those observed experimentally in these regimes. \textbf{Fig. \ref{fig:scale_hblft_tra_ext}(a)} plots the ratio of the flame base lift-off height at extinction ($h_{b,ext}$) with the theoretical estimate of the flame base lift-off height ($h_{b,lft,th}$) estimated at the corresponding $Re_{f}$. Similarly \textbf{Fig. \ref{fig:scale_hblft_tra_ext}(b)} plots the same comparing the extinction time scale ($t_{b,ext}$) with the theoretical estimate of $t_{b,lft}$ ($t_{b,lft,th}$). The plots in \textbf{Fig. \ref{fig:scale_hblft_tra_ext}} also display the error associated with the theoretical prediction based on the data from the re-attachment regimes. Although the experimentally observed values of $h_{b,ext}$ are within the error range associated with theoretical prediction of $h_{b,lft,th}$ (\textbf{Fig. \ref{fig:scale_hblft_tra_ext}(a)}), the timescale associated with extinction ($t_{b,ext}$) is quite lower than the corresponding theoretical predictions (with accounting for the associated error), with an expectation in Regime E-3. In E-3, as the flame base tends to sustain for longer periods of time following tip shedding, the observed values of $t_{b,ext}$ are quite high compared to its theoretical predictions. However, in regimes E-1 and E-2, the flame base lifts off and interacts with either the flame tip (in E-1) or the flame pinch-off point (in E-2) within timescales of $t_{b,ext}$ before reaching a quasi-equilibrium state at timescales of $t_{b,lft,th}$ as observed in the re-attachment regimes. As discussed in the preceding section, this interaction is followed by flame extinction.

    \subsection{Flame Tip undulation and shedding\label{subsec:Flame Tip}} \addvspace{10pt}
    The flame tip tends to exhibit a periodic flickering pattern, which is a result of circulation build-up and associated vortical shedding along the shear boundary surrounding the flame. These shedding behaviours persist even under nominal conditions and tend to become aggravated when subjected to blast waves.
    
\begin{figure*}
        \centering
        \includegraphics[width=0.9\linewidth]{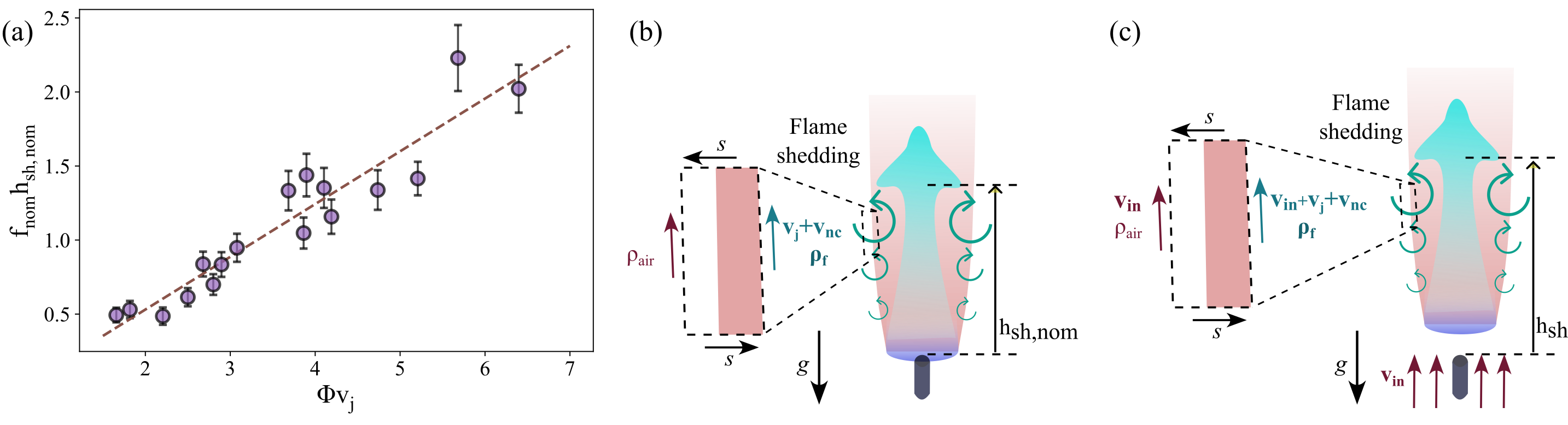}
        \caption{(a) Nominal flickering frequency ($f_{sh,nom}$) expressed as function of fuel jet velocity ($v_{j}$), equivalence ratio ($\Phi$) and flame height ($h_{sh,nom}$). (b,c) Schematic depicting circulation build-up that leads to flame tip shedding in jet flames in quiescent and blast wave imposed flow field conditions, respectively.}
        \label{fig:freq_sch_shed}
    \end{figure*}

    For an open jet flame with a fuel-air jet velocity of $v_{j}$ and equivalence ratio of $\Phi$, in a quiescent environment, the flickering frequency ($f_{sh,nom}$) is found to increase with $v_{j}$ and $\Phi$ and decrease with $h_{sh,nom}$ (\textbf{Fig. \ref{fig:freq_sch_shed}(a)}). Experimental data reveals the following correlation (\textbf{Fig. \ref{fig:freq_sch_shed}(a)}).
    \begin{gather*}
        f_{sh,nom} h_{sh,nom} \sim v_{j} \Phi \tag{7}
    \end{gather*}
    
    A simple analysis extending the vorticity transport equation can be used to understand the observed flickering phenomena. For an axis-symmetry, incompressible open jet flame devoid of swirl, the vorticity transport equation (neglecting viscous dissipation terms) can be integrated over an elemental control volume enclosing the shear boundary surrounding the jet flame (\textbf{Fig. \ref{fig:freq_sch_shed}(b)}). The equation below describes the temporal evolution of circulation within this control volume \cite{xia_vortex-dynamical_2018}.
    \begin{gather*}
        \frac{d\Gamma}{dt}= \rho_{a} g \left(\frac{1}{\rho_f}-\frac{1}{\rho_a}\right) \Delta h+\ \frac{d\Gamma_{ini}}{dt} \tag{8}
    \end{gather*}\label{Main_equation}
    The first term in the above equation represents the circulation induced due to a mismatch in the direction of the density gradient ($\nabla \rho$) across the shear boundary and the pressure gradient imposed by gravity ($\rho g$). The term, $(d\Gamma_{ini}/{dt})$, represents the initial circulation imposed over the system, which can be estimated by integrating the velocity field along the boundary of the control volume.
    As illustrated in \textbf{Fig. \ref{fig:freq_sch_shed}(b)}, in a quiescent environment, the only velocity scale that can contribute to $(d\Gamma_{ini}/{dt})$ is $v_{j} + v_{nc}$, which is present on the flame side of the control side of the control volume.
    \begin{gather*}
        d\Gamma_{ini} = (v_{j} + v_{nc}) dh
    \end{gather*}
    where, $dh = (v_{j} + v_{nc}) dt$
    \begin{gather*}
        \Rightarrow \frac{d\Gamma_{ini}}{dt} = (v_{j} + v_{nc})^2 \tag{9}
    \end{gather*}
    
    Vortices are continuously fed by the shear boundary as they travel downstream, and they tend to detach when their circulation exceeds a critical limit ($\Gamma_{cri}$). If $t_{sh,nom}$ is the time period associated with periodic flame flickering, $\Gamma_{cri}$ can be estimated by integrating $({d\Gamma}/{dt})$ for a period of $t_{sh,nom}$. 
    \begin{gather*}
        \Gamma_{cri} = \int_{0}^{t_{sh,nom}} {\left[\rho_{a} g \left(\frac{1}{\rho_f}-\frac{1}{\rho_a}\right) \Delta h + (v_{j} + v_{nc})^2\right]dt}
    \end{gather*}
    
    \begin{figure*}
        \centering
        \includegraphics[width=0.9\linewidth]{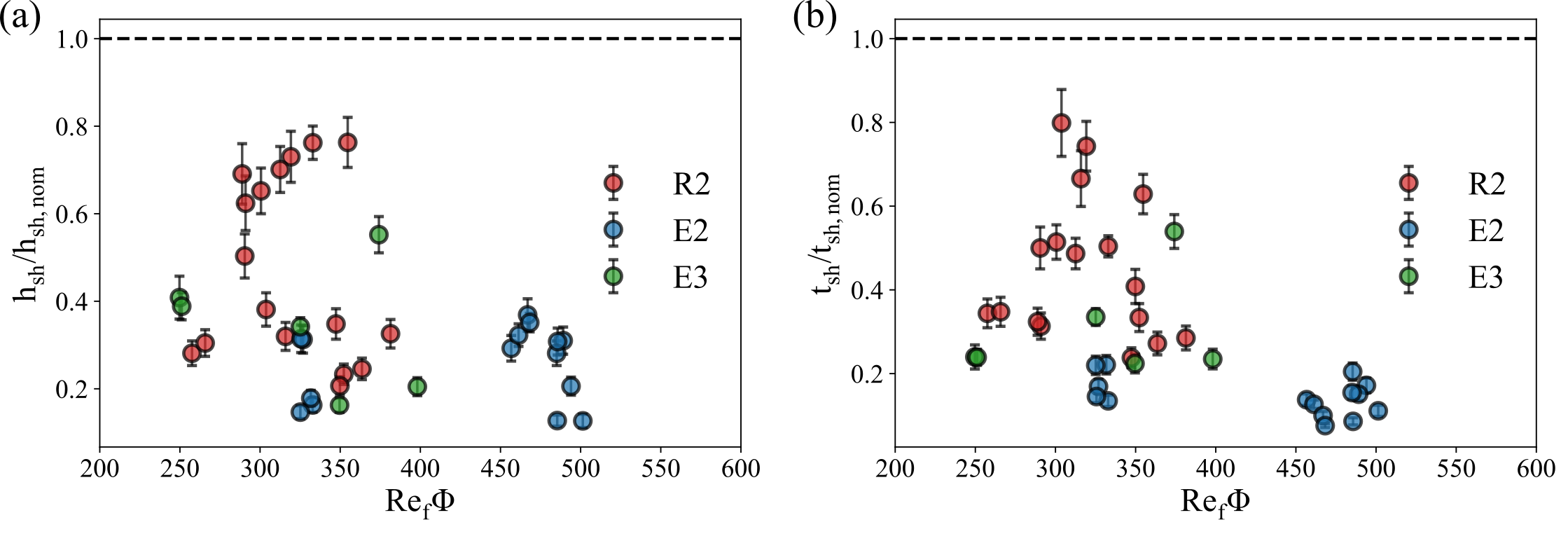}
        \caption{(a,b) Shedding height ($h_{shed}$) and shedding timescales ($t_{shed}$) following the blast wave interaction are compared against corresponding nominal values at the respective fuel-air jet Reynolds number ($Re$) and equivalence ratio ($\Phi$), respectively.}
        \label{fig:shed_f_h}
    \end{figure*}
    
    Following the interaction with the blast wave, the second term in the above integral changes ($(d\Gamma_{ini}/{dt})$ changes) due to the added velocity differential across the shear boundary surrounding the flame. Thus, shear layer vortices tend to reach their critical circulation limit at timescales lower than $t_{sh,nom}$ ($t_{sh}<t_{sh,nom}$). Consequently, this also causes the shedding length scales ($h_{sh}$) to drop below $h_{sh,nom}$ since $h_{sh}$ is essentially the convective length scale associated with the bulk motion of the shear layer vortices in a time scale of $t_{sh}$. As illustrated in \textbf{Fig. \ref{fig:freq_sch_shed}(c)}, $(d\Gamma_{ini}/{dt})$ following the blast wave interaction can be formulated as,
    \begin{gather*}
        \frac{d\Gamma_{ini}}{dt} = \Bigl(v_{j} + v_{nc} + v_{in}\Bigr)^2 - \Bigl(v_{in}\Bigr)^2 \tag{10}
    \end{gather*}
    Thus, as evident from the formulation of $(d\Gamma_{ini}/{dt})$ in the presence of the blast wave (\textbf{Eq. (10)}) and in quiescent conditions (\textbf{Eq. (9)}), an additional term, $2(v_{j} + v_{nc})v_{in}$, is introduced in $({d\Gamma_{ini}}/{dt})$ due to the blast wave, and is responsible for the accelerated circulation build-up in the shear layer vortices. This results in a dip in the timescales and length scales associated with flame tip shedding. \textbf{Fig. \ref{fig:shed_f_h}(a,b)} illustrates the drop in the shedding heights and time scales in comparison with nominal shedding scales under quiescent conditions. 

    \section{Conclusion}
    {
    The work investigates the response characteristics of premixed jet flames following their interaction with blast waves. The study employs a unique facility to generate blast waves, developed based on the wire-explosion technique, and the generated blast fronts are allowed to sweep through the premixed jet flame in the direction of the jet. The flow field (velocity and pressure) imposed by the blast wave on the jet flame can be characterised by a decaying profile following a sharp discontinuity at the blast front. A characteristic feature of the generated blast wave is the bulk flow induced behind it as a result of entrainment from the surroundings following the dip of blast-induced pressure to sub-ambient levels. While the jet flame’s response to the blast front is a jittery motion, the flame base tends to lift off following its interaction with the induced bulk flow. 
    
    The lifted flame is found to further exhibit two major response behaviours: re-attachment and extinction. In the re-attachment regime, the flame base tends to re-attach back at the nozzle tip after attaining a characteristic lift-off height subject to the operating conditions of fuel-air jet Reynolds number, normalised equivalence ratio and incident blast wave Mach numbers. In the extinction regimes, the flame base continues to lift-off and subsequently extinguishes, which is hypothesised to be a result of strain rates exceeding critical limits. Sub-regimes were identified within the re-attachment and extinction regimes based on the flame tip response dynamics following the interaction with the blast front and the induced flow behind it. While the flame tip displayed a mild flame stretching event in the Type-1 sub-regime, flame neck formation followed by a possible flame pinch-off was observed in the Type-2 sub-regime. The formation of a necking zone in the jet flame is a result of vortex roll-up along the shear boundary between the hot product gases and the ambient air surrounding the flame. Flame pinch-off/shedding occurs when these vortices reach their critical circulation limits and shed at length scales lower than the flame height, pinching off a portion of the flame tip. Type-1 and Type-2 flame response behaviours were observed in both re-attachment and extinction regimes. However, an additional sub-regime was identified under the extinction regime, wherein the lifted flame base exhibited intense oscillations in heat release rates and shape following the flame pinch-off event. The flame was observed to eventually extinguish as the flame base lift-off height became comparable to the flame height. A regime map was formulated to identify the parametric space of $Re$, $\phi$ and $M_s$ wherein different flame response behaviours were observed. 
    
    A simplified model extending the co-axial jet approximation was used to explain the observed trends in the flame base response dynamics. A scaling law was formulated for the flame base lift-off height and its associated timescale. The flame tip’s response characteristics were explained by extending the vorticity transport equation to account for vortex roll-up and shedding along the shear boundary surrounding the flame. Observations and results from the study can be used to understand and improve upon existing high-speed propulsion systems such as scram jets and RDEs, wherein flames interact with non-linear pressure waves such as blast waves are common.
    }
    
    \section*{Declaration of Interests}
    {
    The authors report no conflict of interest.
    }
    
    \section*{Acknowledgements}
    {
    The authors are thankful to SERB (Science and Engineering Research Board) - CRG: CRG/2020/000055 for financial support. S.B. acknowledges funding through the Pratt and Whitney Chair Professorship. A.A. acknowledges funding received through the Prime Minister's Research Fellowship scheme.
    }

    \bibliography{mybibfile}

    \end{sloppypar}

    \newpage

    \onecolumn

    \section{Supplementary}
    \addcontentsline{toc}{section}{Supplementary}

    \setcounter{figure}{0}
    \renewcommand{\thefigure}{S\arabic{figure}}

    \section*{Strength of the generated Blast wave at different charging voltages}

    \begin{figure} [hbt!]
        \centering
        \includegraphics [width=0.43\linewidth] {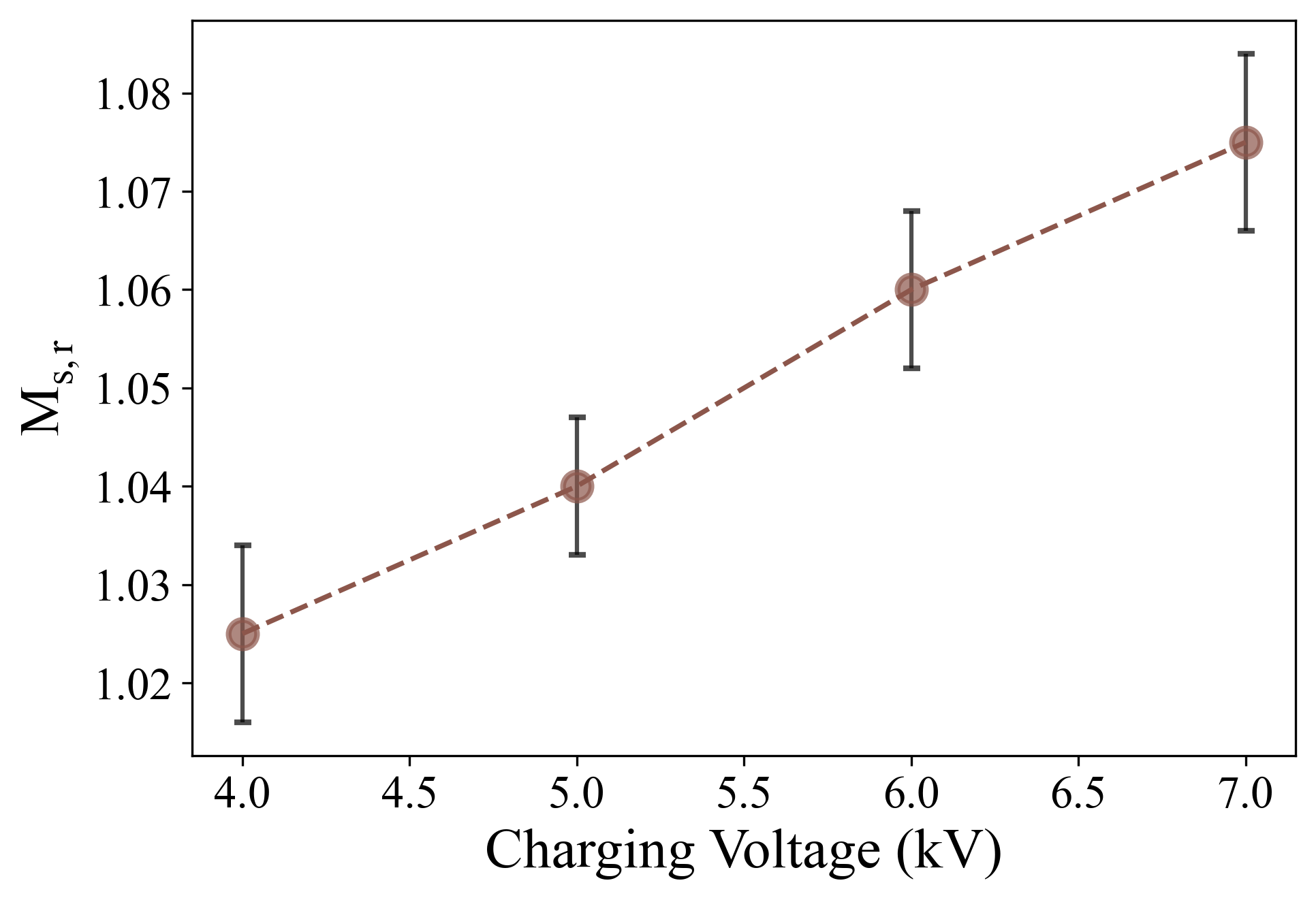}
        \caption{The figure plots the variation of the Mach number of the generated blast front at different charging voltages, as measured at a distance of 264 mm from the source of explosion. The strength of the generated blast wave is found to increase with an increase in the charging voltage.}
    \end{figure}
    
    \section*{Theortical estimates of the timescales associated with the blast wave}
    
    \begin{figure} [hbt!]
        \centering
        \includegraphics [width=0.43\linewidth] {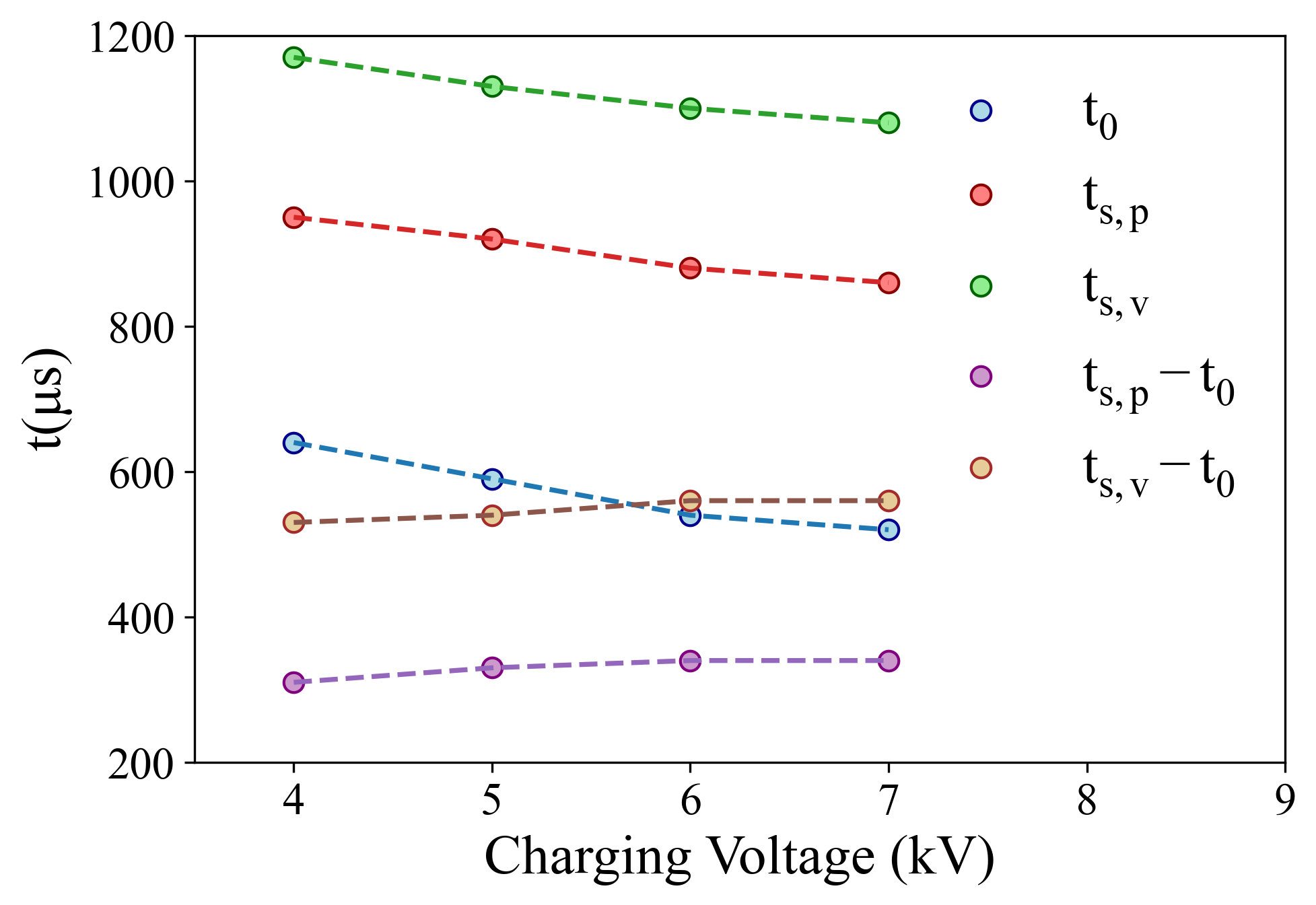}
        \caption{The figure plots the variation of $t_{0}$, $t_{s,p}$, $t_{s,v}$, $t_{s,p} - t_{0}$ and $t_{s,v} - t_{0}$ for different blast strengths (or charging voltage of the blast generation facility). The plots represent the time scales associated with the blast wave imposed velocity and pressure fields as measured from a distance of 264 mm from the source of the explosion.}
    \end{figure} 
    
    \section*{Estimation of induced velocity}
    
    \begin{figure} [hbt!]
        \centering
        \includegraphics [width=0.43\linewidth] {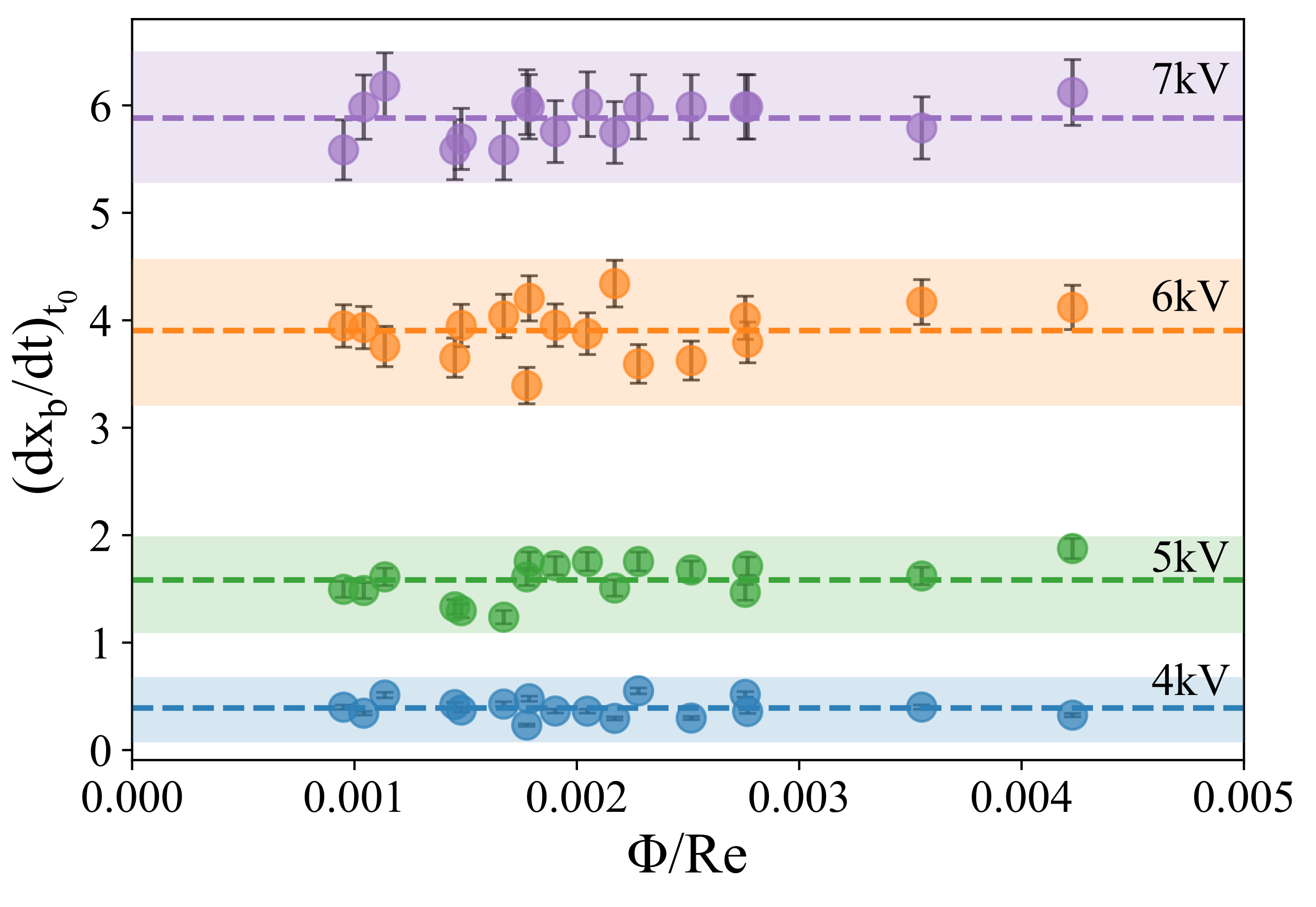}
        \caption{The figure plots the flame base lift-off rate at the instant of the interaction of the premixed jet flame with the flame base. It should be noted that the flame base lift-off rate remains at a near-constant value for a given blast strength, irrespective of the Reynolds number and Equivalence ratio of the premixed flame.}
    \end{figure} 

\end{document}